\documentclass[%
11pt,
 amsmath,amssymb,
 aps,
prfluids
]{revtex4-2}

\usepackage{graphicx}
\usepackage{dcolumn}
\usepackage{bm}

\usepackage[utf8]{inputenc}
\usepackage[T1]{fontenc}
\usepackage{mathptmx}
\usepackage{etoolbox}
\usepackage{yfonts}
\usepackage[dvipsnames]{xcolor}
\usepackage{enumitem}
\usepackage{ulem} 
\usepackage{verbatim} 
\usepackage{siunitx} 
\usepackage{physunits} 
\usepackage[colorlinks=true, allcolors=blue]{hyperref}

\DeclareMathAlphabet\mathbfcal{OMS}{cmsy}{b}{n}

\newcommand*{\xs}{\vec{x}_{\mathrm s}}
\newcommand*{\xsk}{\vec{x}_{{\mathrm s},k}} 
\newcommand*{\xsi}{\vec{x}_{{\mathrm s},i}} 
\newcommand*{\xcm}{\vec{x}_{\mathrm{cm}}}

\newcommand*{\Gg}{\bar{\bar{G}}}
\newcommand*{\Tt}{\bar{\bar{\bar{T}}}}
\newcommand*{\Rr}{\bar{\bar{R}}}

\newcommand*{\Ii}{\bar{\bar{I}}}
\newcommand*{\fint}{\vec{f}^{(\mathrm{int})}}
\newcommand*{\phiint}{\vec{\phi}^{(\mathrm{int})}}

\newcommand*{\Nfext}{N_{\mathrm {f}}^{(\mathrm{ext})}}
\newcommand*{\Nfint}{N_{\mathrm {f}}^{(\mathrm{int})}}
\newcommand*{\utot}{\vec{u}_\mathrm{tot}}

\begin{document}

\preprint{APS/123-QED}

\title{Internal stresses in low-Reynolds-number fractal aggregates}

\author{Matteo Polimeno}
\email{mpolimeno@ucmerced.edu}
\affiliation{%
 Department of Applied Mathematics, University of California, Merced\\
}
\author{Changho Kim}
\affiliation{%
 Department of Applied Mathematics, University of California, Merced\\
}%

\author{Fran\c{c}ois Blanchette}%
\affiliation{%
 Department of Applied Mathematics, University of California, Merced\\
}%




\date{\today}

\begin{abstract}
We present a numerical model of fractal-structured aggregates in low-Reynolds-number flows.
Assuming that aggregates are made of cubic particles, we first use a boundary integral method to compute the stresses acting on the boundary of the aggregates.
From these external stresses, we compute the stresses within the aggregates in order to gain insights on their breakup, or disaggregation.
We focus on systems in which aggregates are either settling under gravity or subjected to a background shear flow and study two types of aggregates, one with fractal dimension slightly less than two and one with fractal dimension slightly above two.
We partition the aggregates into multiple shells based on the distance between the individual cubes in the aggregates and their center of mass and observe the distribution of internal stresses in each shell.
Our findings indicate that large stresses are least likely to occur near the far edges of the aggregates. We also find that, for settling aggregates, the maximum internal stress scales as about 7.5\% of the ratio of an aggregate's apparent weight to the area of the thinnest connection, here a single square.
For aggregates exposed to a shear flow, we find that the maximum internal stress scales roughly quadratically with the aggregate radius.
In addition, after breaking aggregates at the face with the maximum internal stress, we compute the mass distribution of sub-aggregates and observe significant differences between the settling and shear setups for the two types of aggregates, with the low-fractal-dimension aggregates being more likely to split approximately evenly.
Information obtained by our numerical model can be used to develop more refined dynamical models that incorporate disaggregation.
\end{abstract}

\maketitle

\section{\label{sec:Introduction}Introduction}

Near the ocean surface, microorganisms and other particulates tend to cluster into fractal structures as they come into contact with each other~\cite{burd_particle_2009}.
The resulting marine aggregates play an important role in the oceanic carbon cycle~\cite{carbon}.
A similar aggregation and transport mechanism may also be applicable to model microplastics in the ocean, which is a major environmental concern~\cite{microplastics}.
Aggregates initially grow over time and the mechanisms that lead to their formation through random encounters have been studied extensively~\cite{meakin_3D_paper, jung, Polimeno2022}.
However, it is well known that aggregates may also break up, either as they settle under gravity or because of stresses induced by some background flow~\cite{Song}.
A full characterization of the rupture dynamics of aggregates (i.e., disaggregation) is often lacking in numerical models of aggregate dynamics and forms the subject of this paper. 

Recently, there have been renewed efforts to characterize the equilibrium size of aggregates immersed in some background flow, both in experimental settings~\cite{DeLaRosaZambranVerhilleLeGal2018, SongRau2022,BrouzetGuineDalbeFavierVandenbergheVillermauxVerhille2021} and in simulations~\cite{FRUNGIERI2021357, ZhaoVowinckelHsuBaiMeiburg2023}.
For instance, De La Rosa et al.~\cite{DeLaRosaZambranVerhilleLeGal2018} conducted experiments to study the fragmentation of aggregates made of magnetic particles in high-Reynolds-number turbulent von K\'arm\'an flows.
The authors found that given the intensity of turbulence and the cohesive forces and shapes of the individual particles, one can predict the average size of aggregates exposed to turbulent flows.
Zhao et al.~\cite{ZhaoVowinckelHsuBaiMeiburg2023} developed numerical simulations to analyze the flocculation dynamics of aggregates exposed to turbulent shear.
They investigated the influence of the shear rate on aggregate size, and identified what they refer to as ``optimal shear rate'', which is sufficiently large to increase particle concentration, thus promoting aggregate growth, but not so large as to cause breakup.
These studies focused on the impact of strong turbulence acting on large aggregates and the resulting equilibrium aggregate size.
In contrast, we consider here aggregates in low-Reynolds-number flows in the early stages of disaggregation and focus on where breakup may occur.

While there has been significant progress on aggregate modeling in recent years, most numerical models have yet to capture the dynamics of disaggregation in a manner that can be readily incorporated into stochastic particulate models.
Traditionally, disaggregation has been modeled by including a given probability that aggregates might break once they reach a pre-determined critical size~\cite{Kolb_1986}.
In this so-called reversible aggregation approach, aggregates are usually set to break at random locations, regardless of their structure or of the conditions they are exposed to.
To investigate possible breakup mechanisms, Zaccone et al.~\cite{ZacconeSoosLattuadaWuBablerMatthausMorbidelli2009} studied aggregates in the intermediate-Reynolds-number regime and focused on the rupture of compact colloidal aggregates via crack propagation, a mechanism that resembles the fracturing of brittle solid materials.
Their model successfully accounts for the scaling size of colloidal aggregates under an applied hydrodynamic stress.
However, its applicability is limited to systems consisting of compact aggregates with fractal dimensions close to three.
To overcome some of those limitations, Gastaldi and Vanni~\cite{GastaldiVanni2011} proposed a model to characterize the distribution of internal stresses in fractal aggregates based on the method of reflections~\cite{happel1965low}.
In this study, aggregates were made of spheres and settled under the effect of a constant force in an unbounded fluid at rest.
The Stokes equations were approximately solved to compute the flow velocity and extract the drag force acting on each sphere.
These quantities were then used to characterize internal stresses within the aggregates.
In the sample results of computed internal stresses provided, larger internal stresses were systematically found to arise close to the center of the aggregates whereas particles near the edges were observed to experience significantly smaller internal stresses.
While this approach gives accurate results for low-density aggregates, providing a good comparison point for some of our results, no background flows were considered, and the assumption of widely separated particles reduces the accuracy of the method when applied to aggregates in which particles with high coordination number~\cite{German} are present.
The method we present here accurately accounts for the presence of neighboring particles and so relaxes this assumption.

We study here aggregates in low-Reynolds-number flows and present a boundary-integral formulation of the Stokes equations~\cite{pozrikidis_1992} that allows us to characterize the external and internal stresses felt by aggregates subjected to either a gravitational force causing them to settle or a background shear flow.
We study how the internal stresses are distributed within aggregates of two different fractal dimensions and characterize how different conditions might affect the rupture of aggregates.
Moreover, we quantify the magnitude of the largest internal stresses as a function of either the aggregate's apparent weight~\cite{Oman} or the background shear rate.

The rest of this paper is organized as follows.
In Section~\ref{sec:disagg_Methods}, we give the details of the boundary-integral formulation used to compute the stresses on and within aggregates in Stokes flow and explain how we characterize the distribution of internal stresses within an aggregate.
In Section~\ref{sec:disagg_Results}, we present simulation results for aggregates subject either to a constant force or to a shear flow.
We compute the distribution of internal stresses, the scaling of the maximum internal stress, and the relative sizes of aggregates formed after a single breakup event.
Finally, we discuss our results and draw conclusions in Section~\ref{sec:disagg_Conclusion_disagg}.

\section{\label{sec:disagg_Methods}Methods}

\subsection{\label{sec:disagg_aggregationmodel}Types of Aggregates Considered}

\begin{figure}
    \centering
    \includegraphics[width=0.8\linewidth]{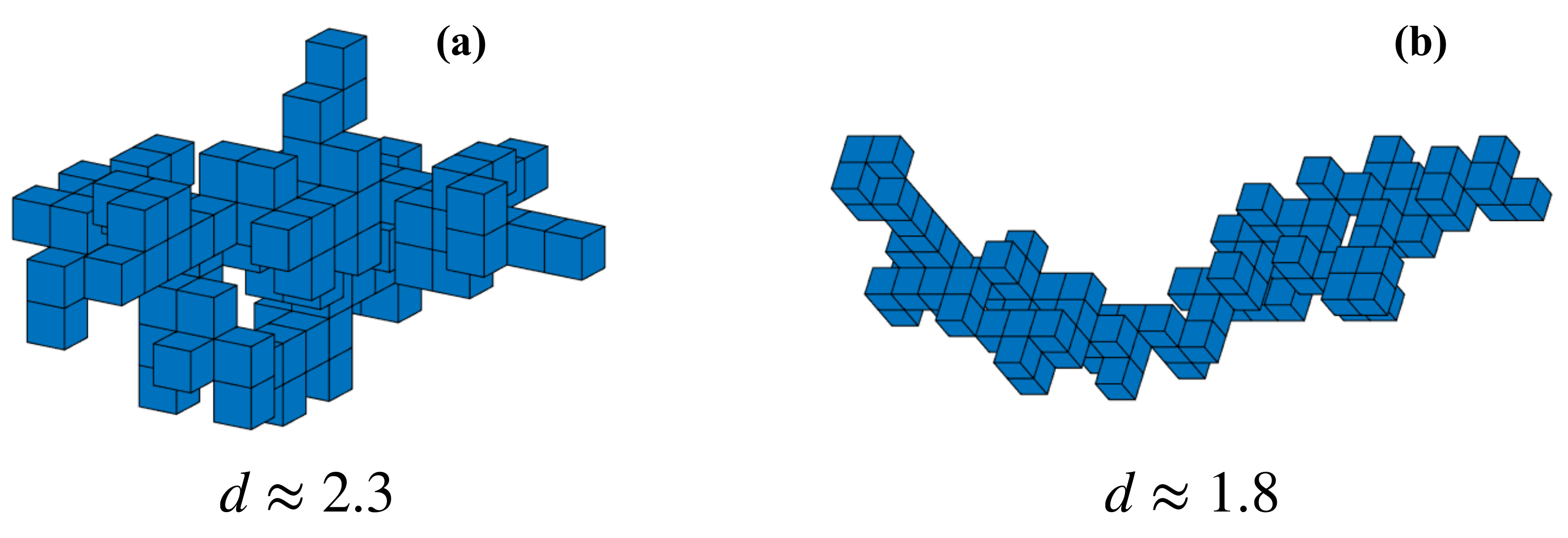}
    \caption{\label{fig:IAAvCCA}
        Panel~(a): typical aggregate formed via IAA-routine; 
        panel~(b): typical aggregate formed via CCA-routine.
        In both cases, aggregates are made of 100 cubes.
        IAA-type aggregates are generally more compact than CCA-type aggregates, as quantified by their fractal dimension $d$ listed under each aggregate.}
\end{figure}

We consider two models of marine aggregates obtained from the well-established numerical framework of Diffusion-Limited Aggregation (DLA)~\cite{witten1981}.
In order to study aggregates of different fractal dimensions~$d$~\cite{meakin1983}, we use two different DLA-based routines to build aggregates: Individually-Added Aggregation (IAA, also known as particle-cluster aggregation) with $d\approx{2.3}$, and Cluster-to-Cluster Aggregation (CCA) with $d\approx{1.8}$.
Our aggregates are built from solid cubic particles on a three-dimensional regular grid, closely following Yoo et al.~\cite{eunji}, to which we refer the interested reader for a detailed description of both the IAA and CCA aggregation routines used in this work.
In Figure~\ref{fig:IAAvCCA}, we display typical aggregates formed by these routines.
As the values of $d$ indicate, an IAA-type aggregate is more compact whereas a CCA-type aggregate is more wispy.
Since our focus is on characterizing the stresses felt by aggregates, we note that one of the main advantages of building aggregates from solid cubes is that we consider stresses on simple squares and avoid singular situations that may occur when other shapes (e.g., spheres) are used.
This fact will be exploited to quickly and accurately compute these stresses via a boundary-integral approach~\cite{pozrikidis_1992}, as discussed in detail in Section~\ref{sec:disagg_externalstresses}.

\subsection{\label{sec:disagg_externalstresses}Computation of External Stresses}

In this paper, we assume that aggregates are made of cubes with side-length $2L=2\unit{\um}$, which approximately corresponds to the diameter of individual phytoplankton cells~\cite{plankton,marinesnow}.
This allows us to neglect inertial effects in the description of the flow around aggregates and model the relevant dynamics simply using Stokes equations,
\begin{equation}\label{eq:stokes}
    \begin{aligned}
        \vec{\nabla}\cdot \utot = 0,\\
        -\vec{\nabla} P_\mathrm{d} + \mu \nabla^2 \utot = 0,
    \end{aligned}    
\end{equation}
where $\vec{u}_\mathrm{tot}$ represents the fluid's velocity and $P_\mathrm{d}$ is the dynamic pressure, defined at a point $\vec{x}_\mathrm{p}$ in the fluid as $P_\mathrm{d}(\vec{x}_\mathrm{p})=P(\vec{x}_\mathrm{p})+\rho{\vec{g}}\cdot\vec{x}_\mathrm{p}$, where $\vec{g}$ is the gravitational acceleration.
In this work, we assume that both the density, $\rho$, and the viscosity, $\mu$, are constant.

To compute the fluid velocity on and around an aggregate in a generic background flow, $\vec{u}^\infty$, we introduce a disturbance flow, caused by the presence of the solid object and define its velocity as $\vec{u}~=~\utot~-~\vec{u}^\infty$.
Here $\vec{u}^\infty$ represents the fluid's velocity in absence of the object and is assumed to satisfy the Stokes equations.
As a consequence of the linearity of this system of equations, the disturbance velocity $\vec{u}$ also satisfies the Stokes equations.
Because it also decays to zero away from the object, we may write the disturbance velocity at any point $\vec{x}_0$ that is external to the surface $S$ of the object using boundary integrals as~\cite{pozrikidis_1992}
\begin{equation}\label{eq:BI}
    \vec{u}(\vec{x}_0) = - \frac{1}{8\pi\mu} \int_{S}\vec{f} (\vec{x})\cdot \Gg(\vec{x},\vec{x}_0)dS(\vec{x}) + \frac{1}{8\pi} \int_{S} \vec{u} (\vec{x})\cdot \Tt(\vec{x},\vec{x}_0)\cdot \hat{n} dS(\vec{x}),
\end{equation}
where
\begin{equation}
    \Gg(\vec{x},\vec{x}_0) = \left( \frac{\Ii}{ \lVert{\vec{x}-\vec{x}_0 \rVert}} + \frac{(\vec{x}-\vec{x}_0)(\vec{x}-\vec{x}_0)}{\lVert{\vec{x}-\vec{x}_0 \rVert}^3}\right),\quad
    \Tt(\vec{x},\vec{x}_0) = -{6}\frac{(\vec{x}-\vec{x}_0)(\vec{x}-\vec{x}_0)(\vec{x}-\vec{x}_0)}{\lVert{\vec{x}-\vec{x}_0\rVert}^5},
\end{equation}
are the so-called single- and double-layer potentials, respectively.
Here $\Ii$ is a three-dimensional identity matrix, $\hat{n}$ is a unit normal to the surface $S$ oriented to point outward, and the norms are Euclidean norms.
Note that \eqref{eq:BI} is also applicable at points that are within or on the surface $S$.
The vector $\vec{f}$ is often referred to as a density function, but in this paper we will take advantage of its physical interpretation and refer to it as the stress vector associated to the disturbance flow.
We are interested in determining the stress vector on the aggregate surface, $\vec{f}_\mathrm{tot}$, which can be obtained from the disturbance stress through $\vec{f}_\mathrm{tot} = \vec{f}^\infty +\vec{f}$, where the contribution to the stress from the flow at infinity can be computed by dotting the stress tensor of the flow at infinity with the unit normal: $\vec{f}^\infty = (-\Ii P^\infty  + \frac{\mu}{2} [\nabla u^\infty + (\nabla u^\infty)^T] ) \cdot \hat{n}$, where $P^\infty$ is the pressure field associated to the flow at infinity.

To solve \eqref{eq:BI} for the stress vector $\vec{f}$, we build on the approach of Yoo et al.~\cite{eunji}, where a novel boundary-integral implementation was first introduced to compute the external stresses felt by low-Reynolds-number aggregates moving with a known translational and angular velocity through a fluid.
Here we instead assume that a given external force and torque are acting on aggregates in the presence of a background flow.
By using a boundary-integral formulation, we then determine the aggregates' translational and angular velocity as well as the external stresses felt by the aggregates.

For ease of notation, in what follows, we will assume that the center of mass of the aggregate is at the origin.
Taking advantage of the no-slip boundary condition on the surface of the solid aggregate, we require that the total velocity of the fluid at any point $\xs$ on the boundary of the aggregate satisfies rigid-body motion,
\begin{equation}\label{eq:solid_body_motion}
    \vec{u}_{\mathrm{tot}}(\xs) = \vec{V}+\vec{\Omega}\times{\xs},
\end{equation}
or, equivalently, $\vec{u}(\xs) = \vec{V}+\vec{\Omega}\times{\xs}-\vec{u}^\infty(\xs)$.
Here, $\vec{V}$ and $\vec{\Omega}$ are the unknown translational and {angular} velocity of the aggregate, respectively. 
To evaluate \eqref{eq:BI} in the limit of an external point $\vec{x}_0$ approaching the surface $S$, we make use of the known theoretical result~\cite{pozrikidis_1992},
\begin{equation}\label{eq:pv}
    \lim_{\vec{x}_0\rightarrow{\xs}} \int_{S}\vec{u}\cdot\Tt(\vec{x},\vec{x}_0)\cdot{\hat{n}}dS(\vec{x}) 
    = 4\pi\vec{u}(\xs)+\int_{S}^\mathrm{PV}\vec{u}(\vec{x})\cdot\Tt(\vec{x},\xs)\cdot\hat{n}dS(\vec{x}),
\end{equation}
where the limit is taken from outside $S$ and the integral on the right-hand side should be interpreted in the principal value (PV) sense.
We thus find
\begin{equation}\label{eq:onehalfu}
    \vec{u}(\xs) = - \frac{1}{8\pi\mu} \int_{S}\vec{f}(\vec{x})\cdot \Gg(\vec{x},\xs)dS(\vec{x}) + \frac{1}{2}\vec{u}(\xs) + \frac{1}{8\pi} \int_{S}^\mathrm{PV}{\vec{u}(\vec{x})}\cdot \Tt(\vec{x},\xs)\cdot \hat{n} dS(\vec{x}).
\end{equation}
Finally, plugging in $\vec{u}(\xs) = \vec{V}+\vec{\Omega}\times{\xs}-\vec{u}^{\infty}(\xs)$ into \eqref{eq:onehalfu}, and using the identities \cite{pozrikidis_1992},
\begin{equation}\label{eq:VandOmega}
    \int_{S}\vec{V}\cdot \Tt(\vec{x},\xs)\cdot \hat{n} dS(\vec{x}) = -4\pi\vec{V},\quad
    \int_{S}(\vec{\Omega}\times{\xs})\cdot \Tt(\vec{x},\xs)\cdot \hat{n} dS(\vec{x}) = -4\pi\vec{\Omega}\times{\xs},
\end{equation}
yields, after simplifications,
\begin{equation}\label{eq:velocity}
    \vec{V}+\vec{\Omega}\times{\xs}
    +\frac{1}{8\pi\mu}\int_{S}\vec{f}(\vec{x})\cdot\Gg(\vec{x},\xs)dS(\vec{x})
    = \frac{1}{2}\vec{u}^\infty(\xs)
    -\frac{1}{8\pi}\int_{S}^\mathrm{PV}\vec{u}^\infty(\vec{x})\cdot\Tt(\vec{x},\xs)\cdot \hat{n} dS(\vec{x}),
\end{equation}
which is a representation formula valid on the surface of solid objects that accounts for the presence of a background flow $\vec{u}^\infty$.
We also note that the identities~\eqref{eq:VandOmega} apply to points on the surface of solid objects, and that the simplification used in {Ref.}~\cite{eunji}, which claims the right-hand side of \eqref{eq:velocity} is identically zero, is only applicable when $\vec{u}^\infty$ is itself a rigid-body motion~\cite{pozrikidis_1992}.

\begin{figure}
    \centering
    \includegraphics[width=0.4\linewidth]{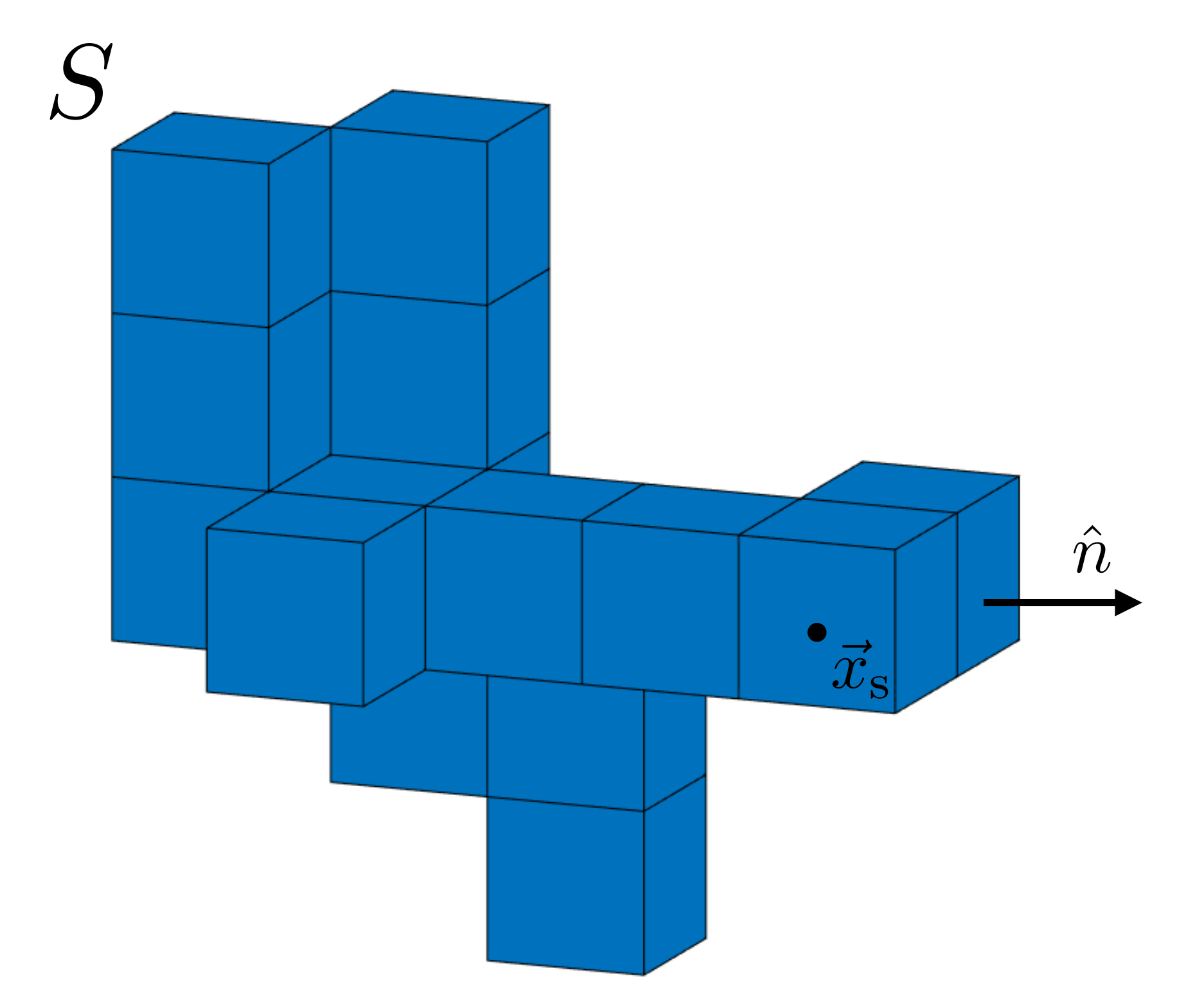}
    \caption{\label{fig:agg_graphics}
        Graphic of how we characterize a given fractal aggregate. Here
        $S$ is the surface of the aggregate, $\xs$ is a point on the surface of the object, and the vector $\hat{n}$ is the outward unit normal to the surface.}
\end{figure}

In addition to the velocity on the surface, we need additional equations to determine the unknown velocity $\vec{V}$ and angular velocity $\vec{\Omega}$.
To this end, we assume that known external force and torque are imposed on the aggregate.
Since all forces must be in equilibrium in inertia-free regimes, we relate the total force, $\vec{F}_\mathrm{tot}$, and torque, $\vec{Q}_\mathrm{tot}$, to the stress vector, $\vec{f}_\mathrm{tot}$, as follows:
\begin{equation}\label{eq:forceandtorquetot}
    \int_S\vec{f}_\mathrm{tot}(\vec{x})dS(\vec{x}) = \vec{F}_\mathrm{tot}, \quad
    \int_S\vec{x}\times\vec{f}_\mathrm{tot}(\vec{x})dS(\vec{x}) = \vec{Q}_\mathrm{tot}.    
\end{equation}
To isolate the disturbance stress $\vec{f}$, we rewrite those equations as
\begin{equation}
    \int_S\vec{f}(\vec{x})dS(\vec{x}) = \vec{F}_\mathrm{tot} -  \int_S\vec{f}^\infty(\vec{x})dS(\vec{x}) = \vec{F}, \quad
    \int_S\vec{x}\times\vec{f}(\vec{x})dS(\vec{x}) = \vec{Q}_\mathrm{tot} -  \int_S\vec{x}\times\vec{f}^\infty(\vec{x})dS(\vec{x}) = \vec{Q}. 
    \label{eq:forceandtorque}
\end{equation}
Combining \eqref{eq:velocity} and \eqref{eq:forceandtorque} yields the following linear system:
\begin{equation}
    \begin{aligned}\label{eq:LS_continous}
        \vec{V}+\vec{\Omega}\times{\xs}
        +\frac{1}{8\pi\mu}\int_S\vec{f}(\vec{x})\cdot\Gg(\vec{x},\xs)dS(\vec{x})
        &= \frac{1}{2}\vec{u}^\infty(\xs)
        -\frac{1}{8\pi}\int_S\vec{u}^\infty(\vec{x})\cdot\Tt(\vec{x},\xs)\cdot\hat{n} \ dS(\vec{x}), \\
        \int_S\vec{f}(\vec{x})dS(\vec{x}) &= \vec{F}, \\
        \int_S\vec{x}\times\vec{f}(\vec{x})dS(\vec{x}) &= \vec{Q}.
    \end{aligned}
\end{equation}
In what follows, we will assume that the stress $\vec{f}$ is constant over each square face, and that the point $\xs$ is always located at the center of a square face.
We provide a simple graphic of how we characterize an aggregate in Figure~\ref{fig:agg_graphics}.

\begin{figure}
    \centering
    \includegraphics[width=0.7\linewidth]{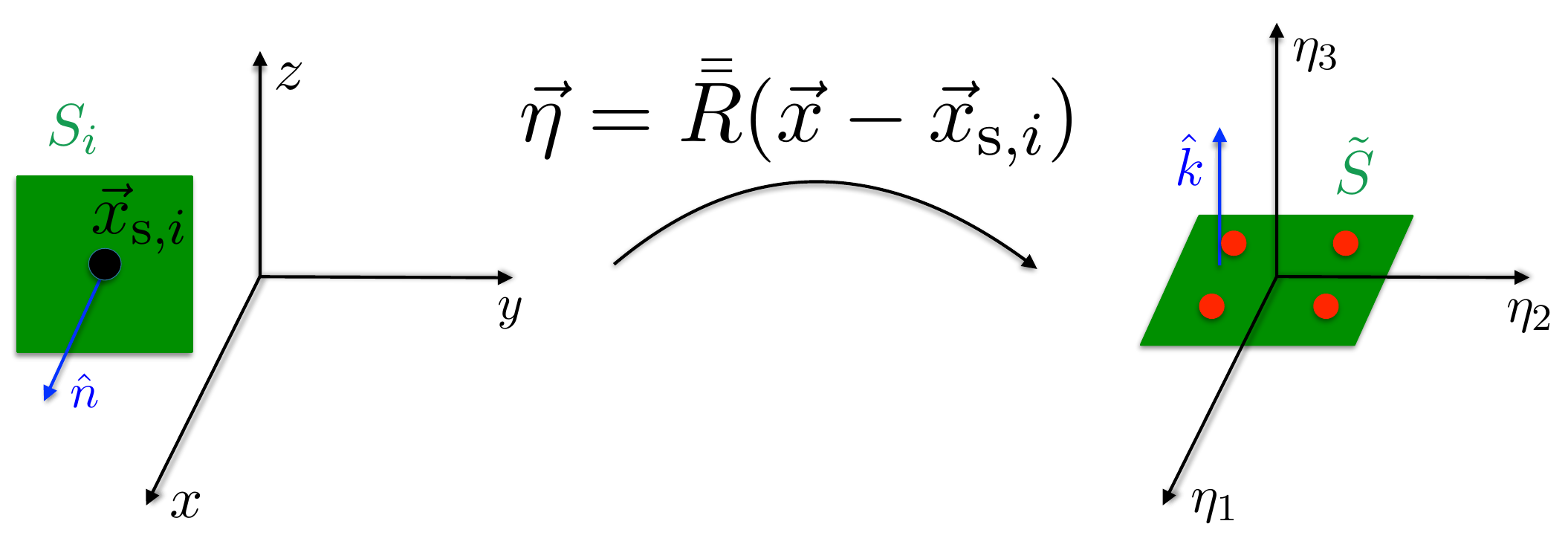}
    \caption{\label{fig:mapping_graphics}
        Graphic of how we map a given face $S_i$ of a cube in an aggregate to the square $\tilde{S}$. 
        The red dots located on $\tilde{S}$ represent the evaluation points of our discretization, for the case in which we were to use only two points in each direction to implement the midpoint rule.}
\end{figure}

To discretize the linear system~\eqref{eq:LS_continous}, we introduce the index $k=1,\dots,\Nfext$, where $\Nfext$ is the number of external faces in the aggregate.
We also introduce another index $i=1,\dots,\Nfext$ that will go over every external square face of the aggregate as the single- and double-layer integrals are computed over the entire boundary $S=\sum_{i=1}^{\Nfext}S_{i}$.
Recall that we evaluate \eqref{eq:velocity} at the center of every square face, $\xsk$, as we vary $k$ to cover every external face one at a time.
For a given $k$, we obtain
\begin{equation}\label{eq:LS_discretized}
    \begin{aligned}
        \vec{V}+\vec{\Omega}\times\xsk+\frac{1}{8\pi\mu}\sum_{i=1}^{\Nfext}\vec{f}_i\cdot\int_{S_i}\Gg(\vec{x},\xsk)dS_i &= \frac{1}{2}\vec{u}^\infty(\xsk) -\frac{1}{8\pi}\sum_{i=1}^{\Nfext}\int_{S_i}\vec{u}^\infty(\vec{x})\cdot
        \Tt(\vec{x},\xsk)\cdot\hat{n}_i dS_i, \\
        \sum_{i=1}^{\Nfext}\vec{f}_i\Delta A &= \vec{F},\\
        -\sum_{i=1}^{\Nfext}\vec{f}_i\times\xsi\Delta A &= \vec{Q},
    \end{aligned}
\end{equation}
where $\Delta A$ is the area of {a} square face.
Once fully discretized, we obtain a dense linear system of size $3(\Nfext + 2) \times 3(\Nfext + 2)$, that can compactly be written as
\begin{equation}\label{eq:linear_system}
    \bar{\bar{A}} \vec{\phi} = \vec{b},
\end{equation}
where
\begin{equation}\label{eq:LS_matrix}
\setlength{\arraycolsep}{3pt} 
\medmuskip = 0.0001mu 
\renewcommand{\arraystretch}{1} 
\bar{\bar{A}}=\left[\begin{array}{cccc|c|c}
\int_{S_1}\Gg(\vec{x},\vec{x}_{{\mathrm s},1})dS_1& \int_{S_2}\Gg(\vec{x},\vec{x}_{{\mathrm s},1})dS_2 &\cdots &\int_{S_{\Nfext}}\Gg(\vec{x},\vec{x}_{{\mathrm s},1})dS_{\Nfext} &\Ii_{1} & [\vec{x}_{{\mathrm s},1}]_{\times}\\
\int_{S_1}\Gg(\vec{x},\vec{x}_{{\mathrm s},2})dS_1& \int_{S_2}\Gg(\vec{x},\vec{x}_{{\mathrm s},2})dS_2 &\cdots &\int_{S_{\Nfext}}\Gg(\vec{x},\vec{x}_{{\mathrm s},2})dS_{\Nfext} &\Ii_{2} & [\vec{x}_{{\mathrm s},2}]_{\times}\\ 
\vdots &\vdots &\ddots &\vdots &\vdots &\vdots\\ 
\int_{S_1}\Gg(\vec{x},\vec{x}_{{\mathrm s},\Nfext})dS_1 & \int_{S_2}\Gg(\vec{x},\vec{x}_{{\mathrm s},\Nfext})dS_2 &\cdots &\int_{S_{\Nfext}}\Gg(\vec{x},\vec{x}_{{\mathrm s},\Nfext})dS_{\Nfext} 
&\Ii_{\Nfext} & [\vec{x}_{{\mathrm s},\Nfext}]_{\times}\\
\hline
\Ii_1\Delta{A} & \Ii_2\Delta{A} & \cdots & \Ii_{\Nfext}\Delta{A} & \bar{\bar{0}} & \bar{\bar{0}}\\
\hline
[\vec{x}_{{\mathrm s},1}]_{\times} & [\vec{x}_{{\mathrm s},2}]_{\times} & \cdots & [\vec{x}_{{\mathrm s},\Nfext}]_{\times} & \bar{\bar{0}} & \bar{\bar{0}}
\end{array}\right],\\
\end{equation}
\begin{equation}
\setlength{\arraycolsep}{0.5pt} 
\medmuskip = 0.0001mu 
\vec{\phi} = 
    \begin{bmatrix}
        \vec{f}_{1}\\
        \vdots \\
        \vec{f}_{\Nfext}\\
        \hline
        \vec{V}\\
        \hline
        \vec{\Omega}\\
    \end{bmatrix},\quad
\vec{b} = \begin{bmatrix}
        ({DL})_{1}\\
        \vdots \\
        ({DL})_{\Nfext}\\
        \hline
        \vec{F}\\
        \hline
        \vec{Q}\\
        \end{bmatrix}.
\end{equation}
Here, we defined the cross-product operator acting on a vector $\vec{z}=[z_1,z_2,z_3]^T$ as
\begin{equation}
    [\vec{z}]_{\times} = 
    \begin{bmatrix}
        0 & -z_3 & z_2\\
        z_3 & 0 & -z_1\\
        -z_2 & z_1 & 0
    \end{bmatrix},
\end{equation}
and
\begin{equation}
    ({DL})_k = \frac{1}{2}\vec{u}^{\infty}(\xsk)-\frac{1}{8\pi}\sum_{i=1}^{\Nfext}\int_{S_i}\vec{u}^{\infty}(\vec{x})\cdot
        \Tt(\vec{x},\xsk)\cdot\hat{n}_i dS_i\quad\mbox{for}\;\; k=1,\dots,\Nfext.
\end{equation}
The single layer integrals are solved analytically as in Ref.~\cite{eunji}, while the double layer integrals that appear in $(DL)_k$ are solved numerically, by first mapping the square $S_i$ to the square $\tilde{S}=\{(\eta_1,\eta_2,\eta_3): -L\le\eta_1\le L,\; -L\le\eta_2\le L,\; \eta_3=0\}$, oriented in the positive vertical direction, see Figure~\ref{fig:mapping_graphics}.
To perform the mapping, we construct a linear operator $\Rr$ such that
\begin{equation}
        \vec{\eta} = \Rr(\vec{x}-\vec{x}_{\mathrm{s},i}),\quad
        \Rr\hat{n} = \hat{k},
\end{equation}
where $\vec{x}_{\mathrm{s},i}$ is the center of $S_i$, $\hat{n}$ is the outward unit normal to $S_{i}$, and $\hat{k}=[0,0,1]^T$ is the unit normal vector in the positive vertical direction.
After the mapping, the double-layer integral becomes
\begin{equation}\label{eq:double_mapped}
    \int_{\tilde{S}}\vec{u}^\infty(\Rr^{T}\vec{\eta}+\vec{x}_{\mathrm{s},i})\cdot
    \Tt(\Rr^{T}\vec{\eta}+\vec{x}_{\mathrm{s},i},\vec{x}_{\mathrm{s},k})\cdot {\Rr^T\hat{k}} d\tilde{S},
\end{equation}
as the Jacobian of the transformation is unity. 
Then, we evaluate \eqref{eq:double_mapped} using the midpoint rule, discretizing $\tilde{S}$ using an even number of points in each direction to avoid evaluation at the singularity.
This allows us to fill the first block of the right-hand side of \eqref{eq:linear_system}.

The second and third blocks on the right-hand side are filled out by the user-input vectors $\vec{F}$ and $\vec{Q}$ for the external disturbance force and torque, respectively.
Note that solving this system yields the translational and angular velocities of the aggregate, $\vec{V}$ and $\vec{\Omega}$, respectively, and the stresses $\vec{f}_k$ felt by the aggregate on its external faces.
We solve \eqref{eq:linear_system} by Gaussian elimination as performed in Matlab.
This system has multiple solutions because a constant pressure, acting in the normal direction, may be added to all external stresses $\vec{f}_k$.
We select the least-squares solution to obtain a unique solution, which corresponds to setting the pressure to zero.
While Gaussian elimination is in general not the most efficient method to solve a large linear system, it was found that, for the typical sizes of aggregate considered in our study, this did not cause any significant additional computational cost.

\subsection{\label{sec:disagg_internalstresses}Computation of Internal Stresses}

To compute the internal stresses, $\fint$, we assume that the total force acting on the aggregate is equally distributed across all its constituting cubes.
This assumption is motivated by a corresponding assumption that the aggregate consists of cubes of constant density, and that the only force other than zero considered in this work is that of gravity.
Therefore, the sum of the stresses over all the faces of a cube times their area equals the total force acting on the aggregate, $\vec{F}_\mathrm{tot}$, divided by the number of cubes in the aggregate,~$M$.
Note that when a cube contains an external face, this sum will involve the previously found external stress on that face $\vec{f}_\mathrm{tot}$.
We further assume that the stresses found on two adjacent faces are equal to each other in magnitude but opposite in direction.
Using these assumptions, we build a linear system $\bar{\bar{C}}\phiint = \vec{d}$ for a vector containing all the internal stresses, $\phiint$, with a $3 \Nfint \times 3\Nfint$ matrix $\bar{\bar{C}}$ and a column vector $\vec{d}$, where $\Nfint$ is the total number of internal faces in the aggregate.
Then, we set up a constrained optimization problem~\cite{bertsekas2014constrained},
\begin{equation}\label{eq:minimization}
\mbox{minimize}\;\;\lVert \phiint \rVert^2 \quad
    \mbox{subject to}\;\;\bar{\bar{C}}\phiint = \vec{d},
\end{equation}
to determine the internal stresses $\phiint$.
We note that the minimization~\eqref{eq:minimization} is required when the constraints do not yield a unique set of internal stresses, as in the situation where four or more cubes form a rectangular prism.
In this scenario, the cubes all share more than one face and precisely chosen internal stresses of any magnitude may cancel each other without affecting the total force on any given cube.

\begin{figure}
    \centering
    \includegraphics[width=\linewidth]{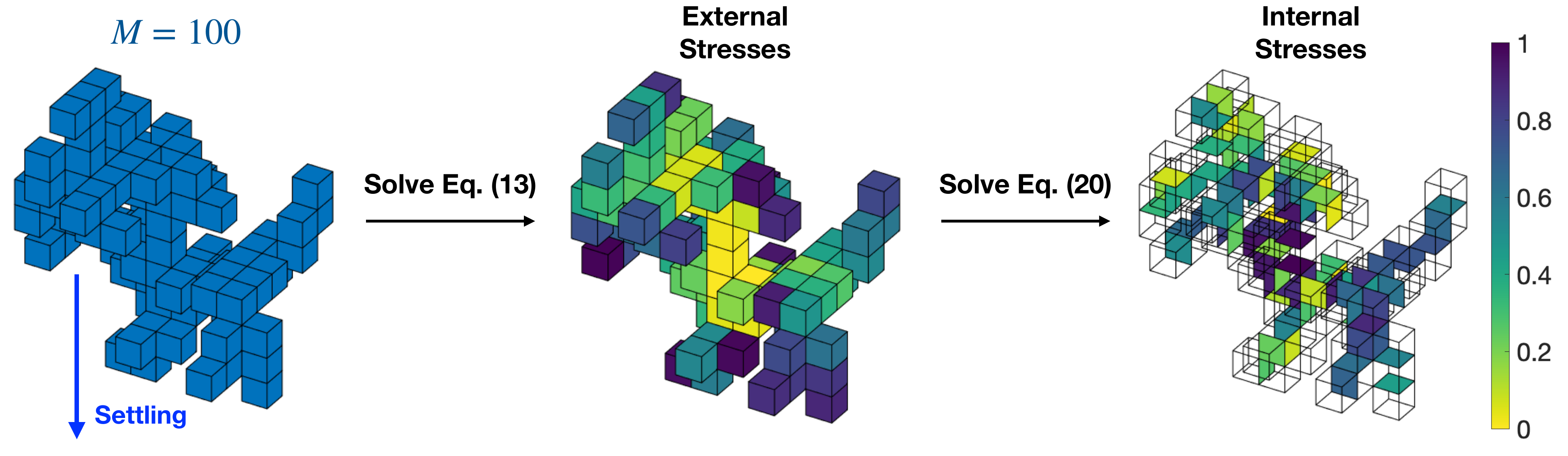}
    \caption{\label{fig:stresses_schematics}
        Simple schematics of how we characterize the stresses in an aggregate made of $M=100$ cubes, settling under gravity. 
        The shape of the aggregate is shown in the leftmost image.
        The middle image depicts the magnitude of all the external stresses found by solving \eqref{eq:linear_system}, normalized by the largest external stress.
        The rightmost image shows the magnitude of all the internal stresses found by solving \eqref{eq:minimization}, normalized by the largest internal stress.        }
\end{figure}

To solve \eqref{eq:minimization}, we first define the Lagrangian function,
\begin{equation}
    \mathcal{L}({\phiint};\vec{\lambda}) = (\phiint)^{T}\phiint+\vec{\lambda}^{T}(\bar{\bar{C}}\phiint-\vec{d}),
\end{equation}
where $\vec{\lambda}$ is a column vector of Lagrange multipliers.
Then we obtain the optimality conditions,
\begin{equation}\label{eq:optimality}
    \begin{aligned}
        \vec{\nabla}_{\phiint}\mathcal{L} & = 2\phiint +\bar{\bar{C}}^{T}\vec{\lambda} = \vec{0},\\
        \vec{\nabla}_{\vec{\lambda}}\mathcal{L} & = \bar{\bar{C}}\phiint-\vec{d} = \vec{0},
    \end{aligned}
\end{equation}
whose solution gives the internal stresses, $\phiint$, that satisfies \eqref{eq:minimization}.
Once the stress vectors on every internal face have been computed, we use their magnitude as a proxy for the likelihood that an aggregate may break.
A simple schematics of the {overall computational procedure} for a typical aggregate made of 100 cubes settling under gravity is shown in Figure~\ref{fig:stresses_schematics}.

\subsection{\label{sec:disagg_shells}Characterizing the Distribution of the Internal Stresses}

\begin{figure}
    \centering
    \includegraphics[width=0.8\linewidth]{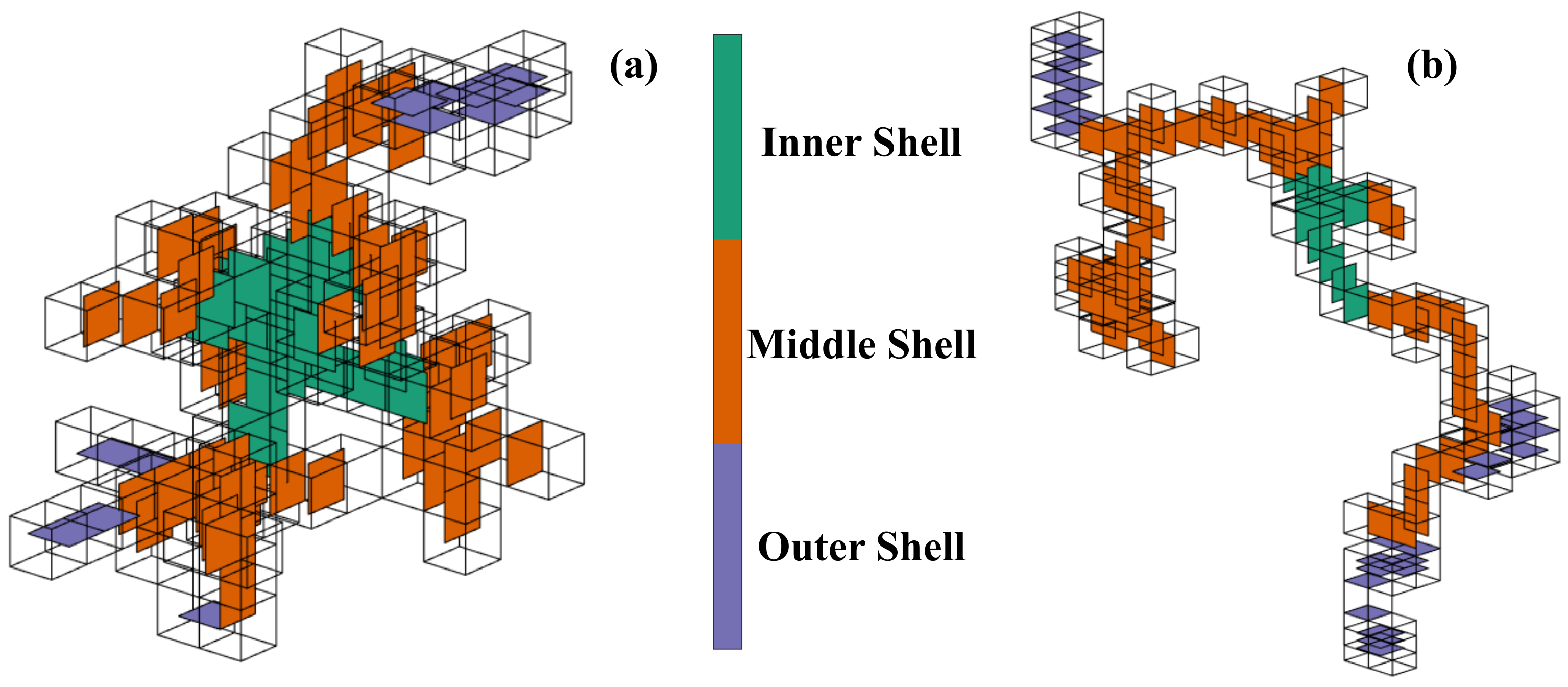}
    \caption{\label{fig:shells}
        Schematics of how we characterize the shells in an aggregate made of $M=100$ cubes, for IAA-type, panel~(a), and CCA-type aggregates, panel~(b).
        The internal faces of the aggregates belonging to the inner, middle, and outer shells are depicted in green, orange, and purple, respectively.
        All the external faces are made transparent for visual clarity.}
\end{figure}

To quantify how the internal stresses distribute across an aggregate's structure, we compute the distance, $R$, between the internal faces of the aggregate and its center of mass, $\xcm$.
Then, letting $\vec{x}_m$ be the center of the $m$-th cube in an aggregate made of $M$ cubes, we define the maximum radius,
\begin{equation}\label{eq:Rmax}
    R_{\mathrm{max}} = L + \max_{m=1,\dots,M}\lVert{\vec{x}_m-\xcm}\rVert,
\end{equation}
and partition the aggregates into three regions, which we refer to
\begin{itemize}[noitemsep,topsep=0pt]
    \item Inner shell: $R < \frac13 R_\mathrm{max}$,
    \item Middle shell: $\frac13 R_\mathrm{max} \leq R < \frac23 R_\mathrm{max}$,
    \item Outer shell: $R\geq \frac23 R_\mathrm{max}$.
\end{itemize}
While we have studied other possible subdivisions of the aggregates, we found that for the typical sizes analyzed in this work, these three shells provide a good quantitative picture of the distribution of the internal stresses.
We provide a simple graphic of the shells for a typical aggregate made of $M=100$ cubes in Figure~\ref{fig:shells}.
To characterize the distribution of internal stresses, we compute histograms for the magnitude of internal stress within each shell.

\begin{figure}
    \centering
    \includegraphics[width=\linewidth]{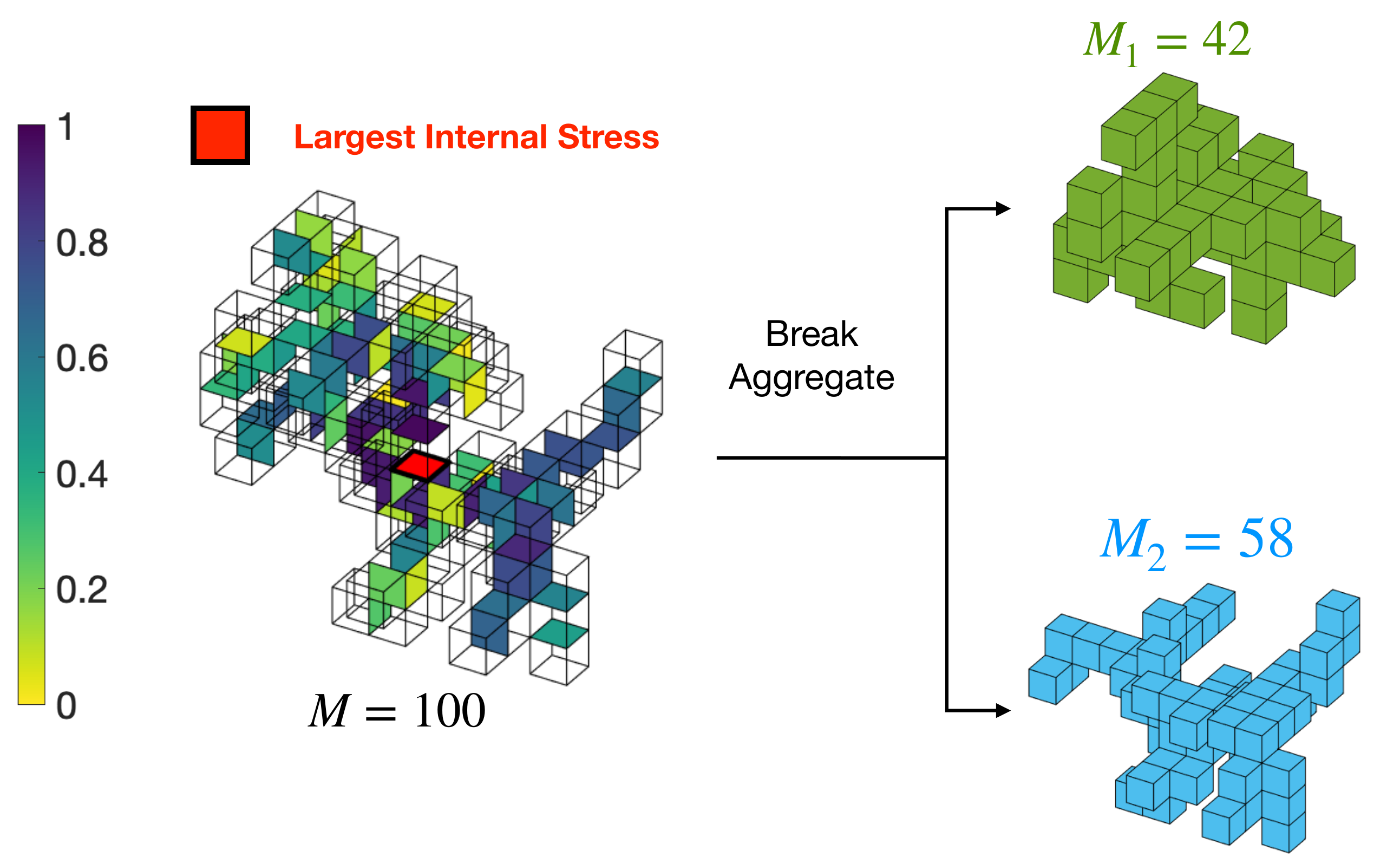}
    \caption{\label{fig:disaggregation_schematics}
        Schematics of our disaggregation routine.
        On the left, we show the distribution of the internal stresses found by solving \eqref{eq:minimization}, and normalized by the largest internal stress (the location of which is highlighted in red), in an aggregate made of $M=100$ cubes, for the Settling Case.
        On the right, we show the two sub-aggregates resulting from the breakup.}
\end{figure}

To further gain insight on disaggregation modeling, we break the aggregate in two at the location of the internal face where the maximum internal stress is found, as depicted in Figure~\ref{fig:disaggregation_schematics} for a typical aggregate made of $M=100$ cubes, settling under gravity. 
We label the sub-aggregate with the smaller number of cubes with subscript~1, and the one with the greater number of cubes with subscript~2.
For each sub-aggregate obtained, we compute the number of their constituting cubes, to quantify the size and mass of the two sub-aggregates compared to the original aggregate.

\subsection{\label{sec:disagg_Cases}Cases Considered}

We aim to characterize the internal stresses induced on the aggregates either by the action of some external force, or by some background flow.
To this end, we study the following two cases:
\begin{itemize}[topsep=0pt,noitemsep]
    \item Settling in a fluid at rest (\textbf{Settling Case}): $\vec{u}^\infty(\vec{x}) = \vec{0}$, $\vec{f}^\infty(\vec{x}) = \vec{0}$, $\vec{F}_\mathrm{tot} = (\rho_\mathrm{agg}-\rho)\vec{g}MV$, $\vec{Q}_\mathrm{tot}=\vec{0}$; 
    \item Force-free shear flow (\textbf{Shear Case}): $\vec{u}^\infty(\vec{x}) = [\gamma_t y,0,0]^T$, $\vec{f}^\infty(\vec{x}) = \frac12 \mu\gamma_t [n_y,n_x,0]^T$, $\vec{F}_\mathrm{tot} = \vec{0}$, $\vec{Q}_\mathrm{tot}=\vec{0}$, where $n_x$ and $n_y$ are the $x$ and $y$ components of the unit normal to the surface of interest, respectively.
\end{itemize}
In the Settling Case, we impose an external force $\vec{F}_\mathrm{tot}$ given by the apparent weight of the aggregate and therefore proportional to the number of cubes constituting a given aggregate, $M$, while we set the background flow $\vec{u}^\infty(\vec{x})$, the background stress $\vec{f}^\infty$, and the external torque $\vec{Q}_\mathrm{tot}$ to zero.
Here, $\rho$ is the density of the fluid, $\rho_\mathrm{agg}$ is the density of the aggregate, $\vec{g}$ is the acceleration due to gravity, and $V$ is the volume of a single cube in the aggregate.
In the Shear Case, we impose a canonical laminar shear flow~\cite{Song} in the horizontal direction with shear rate $\gamma_t$ and constant associated stress.
For this case, we set both the external force $\vec{F}_\mathrm{tot}$ and torque $\vec{Q}_\mathrm{tot}$ to zero, and assume that the center of mass of the aggregate is located at the origin.

We remark that in both of those cases, the disturbance force and torque are actually equal to the total force and torque. 
Moreover, because the system is linear, all computed stresses are proportional to either the imposed force or the shear rate.
In the Shear Case, stresses are also proportional to the viscosity, while in the Settling Case the viscosity is inversely proportional to the velocity and angular velocity but does not affect the computed stresses.
For both cases, we compute the internal stresses for aggregates made of a range of $M$ cubes ($M=25,50,100,150,200$), collecting 400 samples for each value of $M$. 
For the Shear Case, to evaluate \eqref{eq:double_mapped} using the midpoint rule, we discretize the square face $\tilde{S}$ using 10 points in each direction.
To analyze the distribution of the internal stresses relative to the size and structure of the aggregates, we investigate the distribution of internal stresses using the three shells described in Section~\ref{sec:disagg_shells}.

\section{\label{sec:disagg_Results}Results}

We compare results obtained for the Settling and Shear Cases for both IAA- and CCA-type aggregates.
Our main aim is to characterize the impact of aggregate structure on the distribution of the internal stresses and understand how different external conditions affect the rupture of aggregates.
We report internal stresses rescaled by an appropriate reference stress chosen for each case under study.
For the Settling Case, the reference stress used is the ratio of the force acting on the aggregate, its apparent weight $(\rho_\mathrm{agg}-\rho){g}MV$, to the area of the face of a cube, $(2L)^2$, which yields the rescaled stress
\begin{equation}\label{eq:fintset}
    \vec{f}^{(\mathrm{int})}_\mathrm{settl} = \frac{\vec{f}^{(\mathrm{int})} (2L)^2}{(\rho_\mathrm{agg}-\rho)gMV} = \frac{\vec{f}^{(\mathrm{int})}}{2L(\rho_\mathrm{agg}-\rho)gM}.
\end{equation}
For the Shear Case, the reference stress is the viscous stress given by the product of the shear rate, $\gamma_t$, and the fluid viscosity, $\mu$, which yields
\begin{equation}\label{eq:fintshe}
    \vec{f}^{(\mathrm{int})}_\mathrm{shear} = \frac{\vec{f}^{(\mathrm{int})}}{\gamma_t \mu}.
\end{equation}
We also report the relative mass distributions of aggregates after breakup, using the internal face with  maximal internal stress as a breakup location.
We note that for both the Settling and Shear Cases there are rare instances (less than 5\% of the total), in which the maximum internal stress is found at faces where severing the bond does not result in breakup because of a looped structure or adjacent connected faces. 
In such cases we do not collect any data.
The results for the Settling Case are presented in Section~\ref{sec:disagg_settling} and those for the Shear Case are presented in Section~\ref{sec:disagg_shear}.

\subsection{Settling Case}\label{sec:disagg_settling}

\begin{figure}
    \centering
    \includegraphics[width=\linewidth]{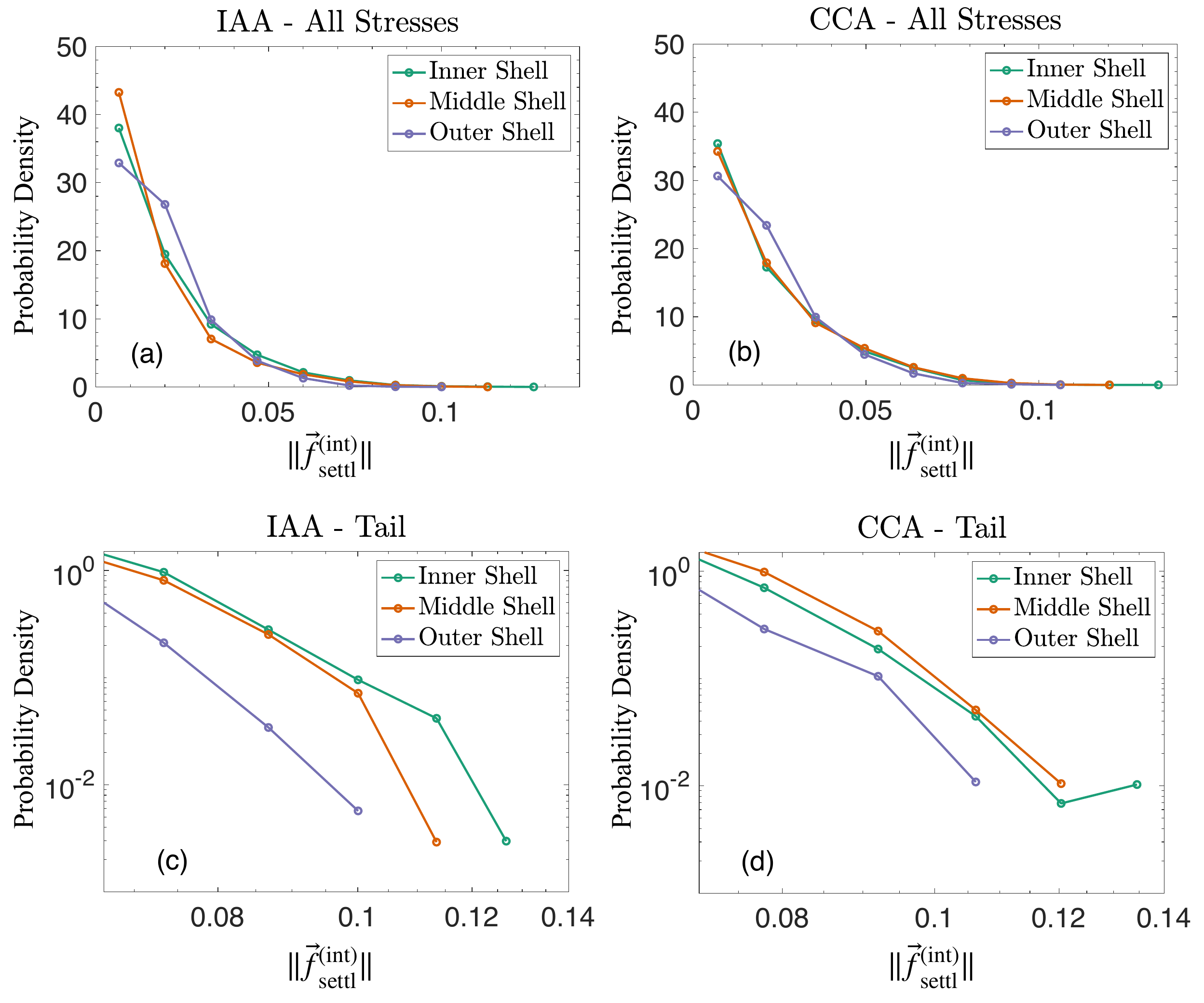}
    \caption{\label{fig:trans_shells}
        Distribution of the magnitude of the rescaled internal stresses for the Settling Case, $\vec{f}^{(\mathrm{int})}_\mathrm{settl}$, see \eqref{eq:fintset}, in the inner (green),  middle  (orange), and outer (purple) shells for IAA-type, panels~(a) and (c), and CCA-type aggregates, panels~(b) and (d).
        In panels~(a) and (b) we show the full range of inner stresses found in the three shells, while in panels~(c) and (d) we zoom in on the tail of the distribution and display it on a log-log scale only for the range greater than half of the maximum internal stress for IAA-type and CCA-type aggregates, respectively.}
\end{figure}

We begin by studying the Settling Case, corresponding to an aggregate denser than the ambient, settling under gravity in a fluid at rest.
In Figure~\ref{fig:trans_shells}, we show the distribution of the magnitude of the rescaled internal stresses $\vec{f}^{(\mathrm{int})}_\mathrm{settl}$, see \eqref{eq:fintset}, in the three shells discussed in Section~\ref{sec:disagg_shells}, computed using 400 samples of a single aggregate made of $M=200$ cubes.
In panels~(a) and (b), we display the distributions of internal stresses computed for IAA-type and CCA-type aggregates, respectively, for the entire range of internal stresses.
To look more closely at the distribution tails, we plot in panels~(c) and (d) these distributions on a $\log$-$\log$ scale focusing only on stress values greater than half the maximum internal stress found across samples for IAA-type and CCA-type aggregates, respectively.
For all shells and for both cases, we observe that the corresponding probability density function decays monotonically, with small stresses occurring most frequently, and that the decay of the tail is consistent with a power-law-like behavior.
Furthermore, we find that approximately 98\% of the stresses have a magnitude smaller than half the maximum stress observed, leaving only rare instances of high stress.
In IAA-type aggregates, the inner shell has roughly 2\% of stresses that lie above this threshold, the middle shell has 1.5\% of such stresses, and the outer shell less than 1\%.
For CCA-type aggregates, the inner shell has 1.4\% of large stresses, while the middle and outer shells have, respectively, 2\% and less than 1\% of large stresses.
We also see that for both types of structures analyzed, the largest stresses always arise in the inner and middle shells.
This indicates that settling IAA- and CCA-type aggregates are more likely to experience stresses that can lead to rupture away from the periphery of their structure rather than nears its edges.

\begin{figure}
    \centering
    \includegraphics[width=0.45\linewidth]{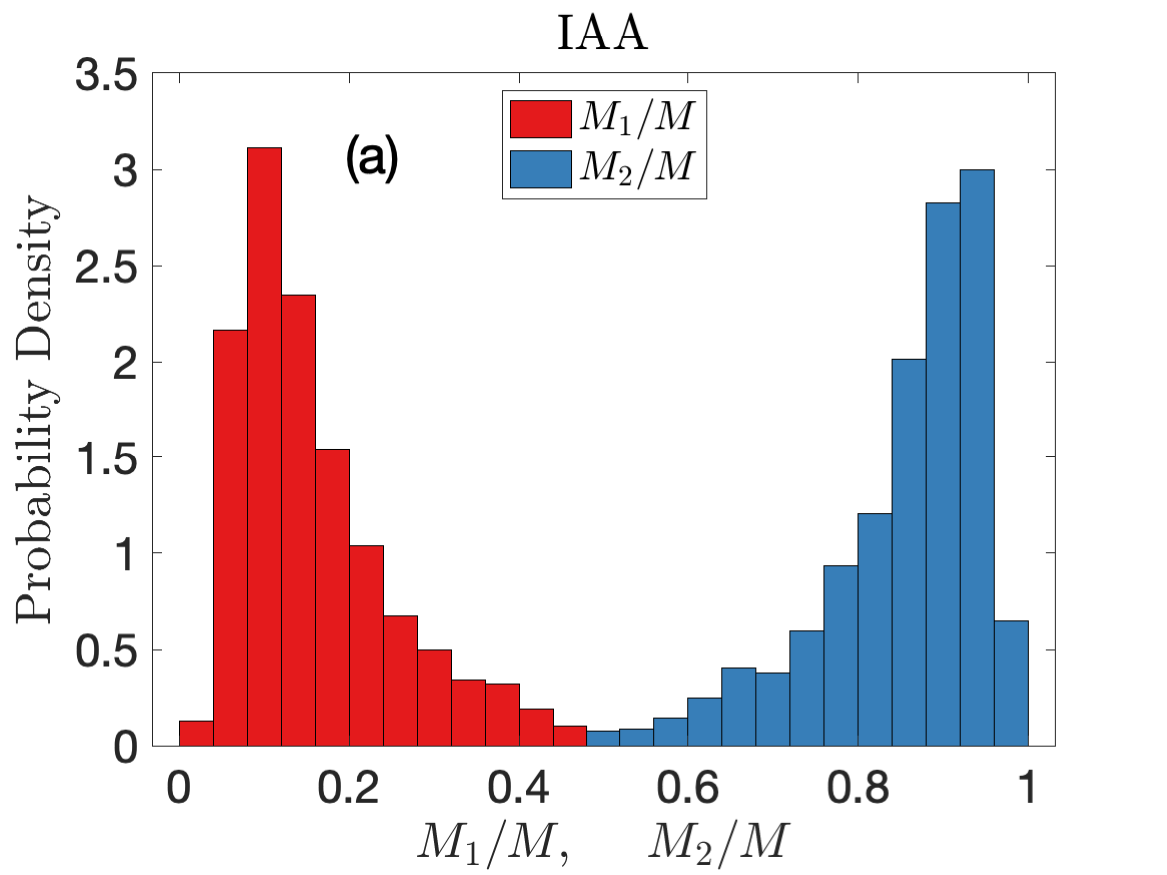}
    \includegraphics[width=0.45\linewidth]{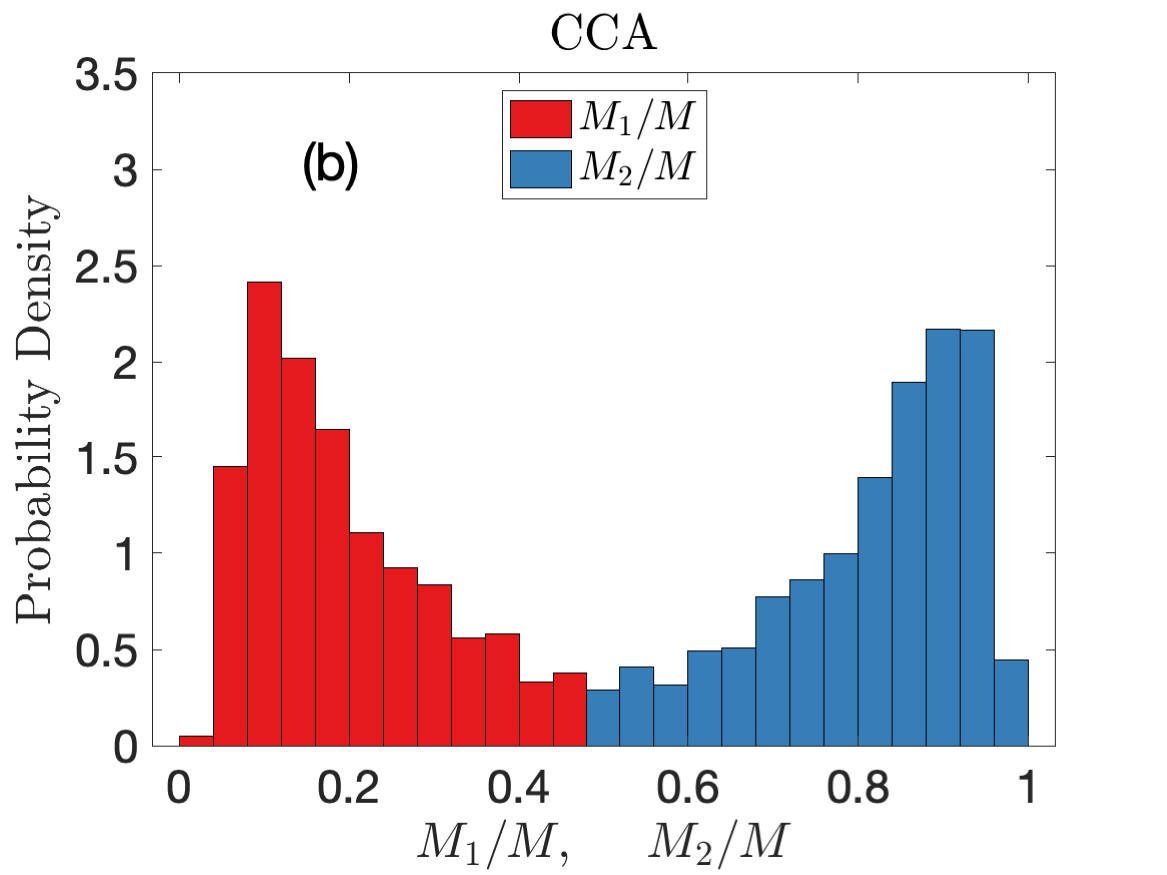}
    \caption{\label{fig:breakup_settling}
        Distribution of the relative masses of post-rupture aggregates for IAA-type, panel~(a), and CCA-type aggregates, panel~(b).
        Samples collected for aggregates with all reported $M$ values were used.
        We display the distribution of the ratio between the mass of the smaller of the two aggregates and the original, $M_1/M$, in red and that of the other ratio $M_2/M$ in blue.
        Results from aggregates that did not break up into two sub-aggregates were discarded.}
\end{figure}

We next analyze the relative masses of the two aggregates formed after rupture, compared to the mass of the original aggregate.
To do so, we use as a breakup location the face where the maximum internal stress occurs.
Figure~\ref{fig:breakup_settling} shows the relative mass distributions of the resulting sub-aggregates, $M_1/M$ and $M_2/M$.
For both IAA-type and CCA-type aggregates, the peaks of the $M_1/M$ and $M_2/M$ distributions are located far from 0.5 (corresponding to $M_1=M_2=M/2$).
This implies that rupture is likely to lead to sub-aggregates with significantly uneven masses.
By comparing the distributions of IAA-type and CCA-type aggregates, one can also see that CCA-type aggregates have a slightly higher tendency of breaking into two aggregates with roughly equal masses, compared to IAA-type aggregates, which appear to break more frequently into two aggregates with uneven masses.

\begin{figure}
    \centering
    \includegraphics[width=\linewidth]{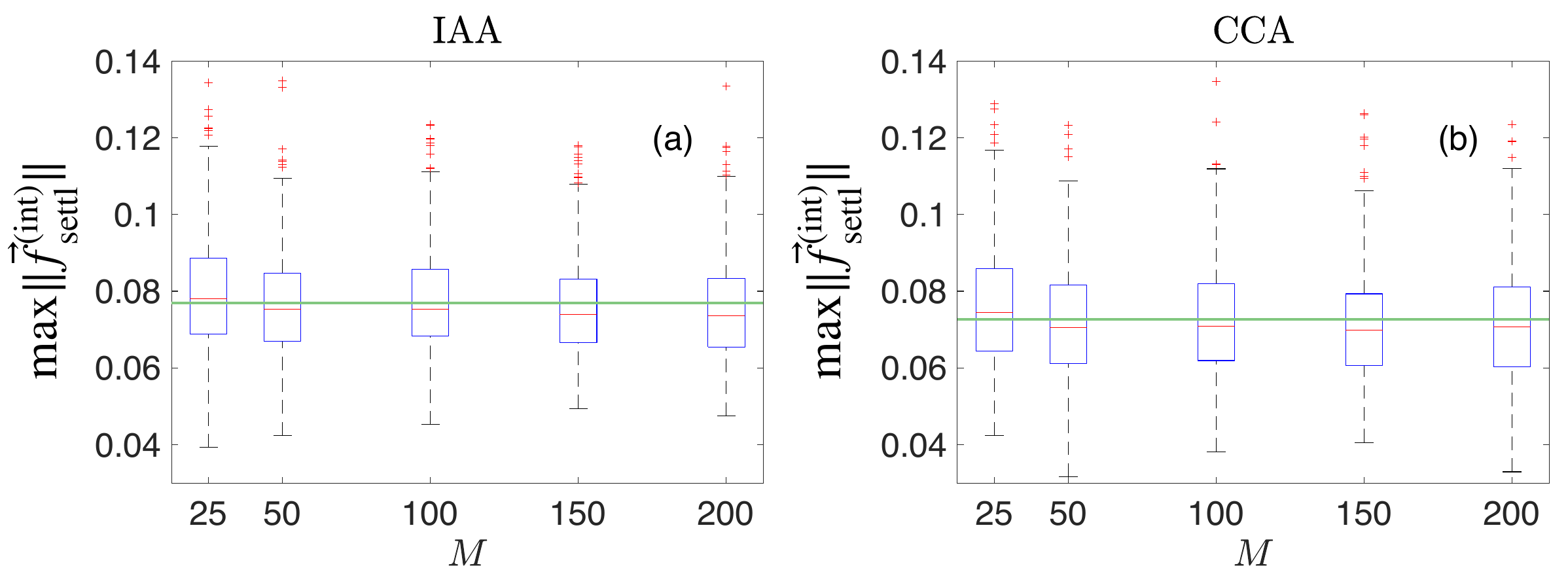}
    \caption{\label{fig:max_stress_vs_M_settling}
        Box and whisker plots of the maximum magnitude of the rescaled internal stress, $\vec{f}^{(\mathrm{int})}_\mathrm{settl}$, see \eqref{eq:fintset}, for IAA-type aggregates, panel~(a), and CCA-type aggregates, panel~(b), for the Settling Case. 
        The solid lines represent the best constant fits to the data.}
\end{figure}

The external force imposed in the Settling Case is the apparent weight of an aggregates, $(\rho_\mathrm{agg}-\rho)gMV$.
To verify whether the maximum internal stress, $\max\left\lVert\vec{f}^{(\mathrm{int})}\right\rVert$, also scales linearly with $M$ as the apparent weight, we look at the maximum magnitude of rescaled stress, $\max\left\lVert\vec{f}^{(\mathrm{int})}_\mathrm{settl}\right\rVert$, for various values of~$M$; for the relation between $\vec{f}^{(\mathrm{int})}$ and $\vec{f}^{(\mathrm{int})}_\mathrm{settl}$, see \eqref{eq:fintset}.
In Figure~\ref{fig:max_stress_vs_M_settling}, we display box and whisker plots of $\max\left\lVert\vec{f}^{(\mathrm{int})}_\mathrm{settl}\right\rVert$ for different values of $M$ for IAA-type, panel (a), and CCA-type, panel (b), aggregates.
For both types, our results consistently show that the maximum magnitude of rescaled stress is independent of $M$, implying that the maximum internal stress corresponds to a constant fraction of the apparent weight divided by the area of a single square, see~\eqref{eq:maxf_settl}.
The solid lines in the figure represent the best constant fits to the data with the constant values of $\alpha=0.077$ for IAA and $\alpha = 0.073$ for CCA.

\subsection{Shear Case}\label{sec:disagg_shear}

\begin{figure}
    \centering
    \includegraphics[width=\linewidth]{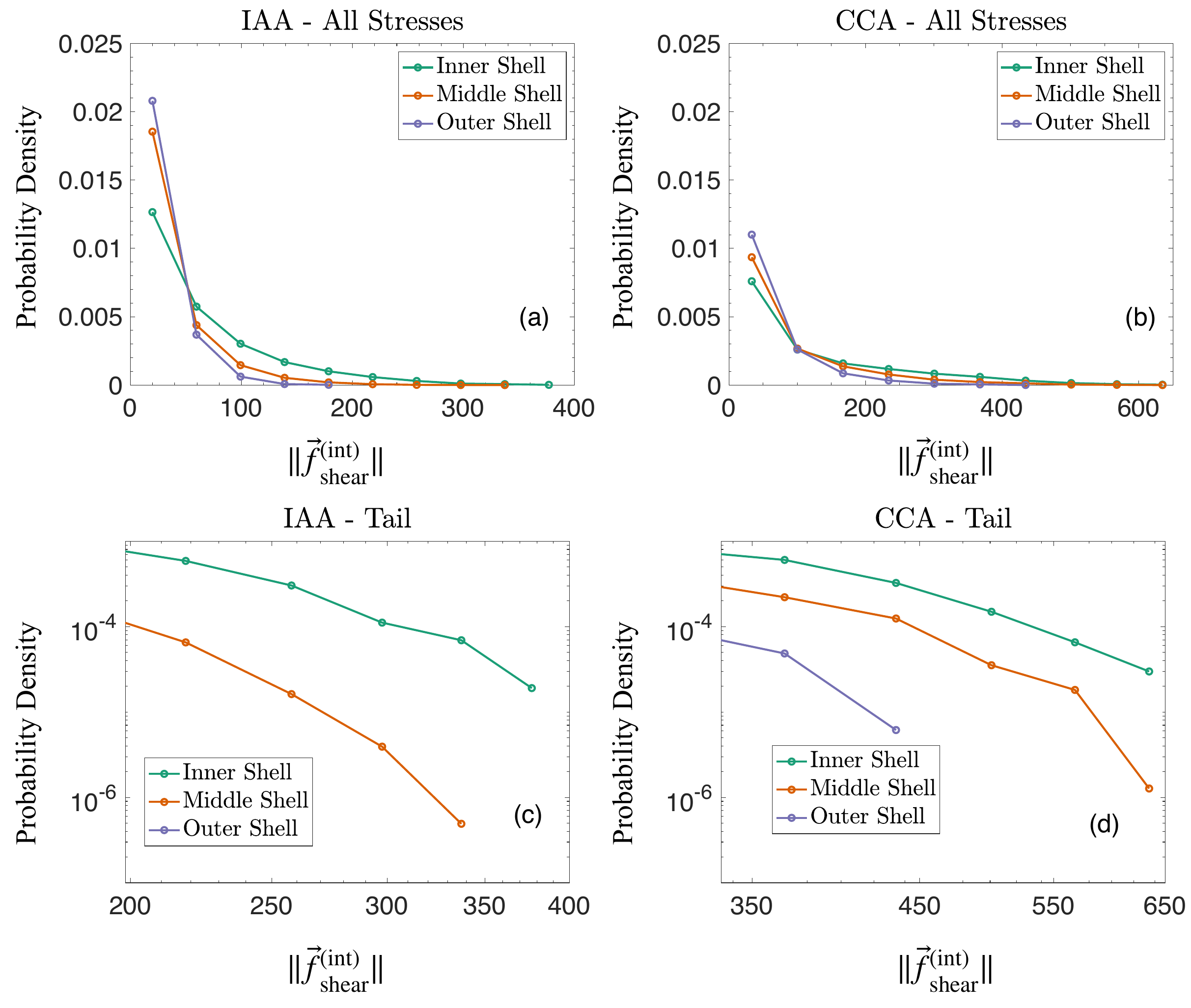}
    \caption{\label{fig:shear_shells}
        Distribution of the magnitude of rescaled internal stresses for the Shear Case, $\vec{f}^{(\mathrm{int})}_{\mathrm{shear}}$, see \eqref{eq:fintshe}, in the inner (green),  middle  (orange), and outer (purple) shells for IAA-type, panels~(a) and (c), and CCA-type aggregates, panels~(b) and (d). 
        In panels~(a) and (b) we show the full range of inner stresses found in the three shells, while in panels~(c) and (d) we zoom in on the tail of the distribution and display it on a log-log scale only for the range greater than half of the maximum internal stress computed for IAA-type and CCA-type aggregates, respectively.
        Note that the outer shell result is not shown in panel~(c) because no stresses larger than half the maximum stress were observed in the outer shell for IAA-type aggregates.}
\end{figure}

We now consider neutrally buoyant aggregates subject to no force or torque in a local shear, which we use as a model of turbulent background flow.
In Figure~\ref{fig:shear_shells} we show the distribution of the magnitude of the rescaled internal stresses, $\vec{f}^{(\mathrm{int})}_\mathrm{shear}$, see \eqref{eq:fintshe}, in the three shells discussed in Section~\ref{sec:disagg_shells}, {computed using} 400 samples for IAA-type and CCA-type aggregates made of $M=200$ cubes.
In panels~(a) and (b) we display the total range of stresses computed for IAA-type and CCA-type aggregates, whereas in panels~(c) and (d) we plot these distributions on a $\log$-$\log$ scale focusing only on the range greater than half the maximum internal stress found across samples for IAA-type and CCA-type aggregates, respectively.
As in the Settling Case, the overall trend is similar for both types of structures, with smaller stresses being more frequent, and a greater proportion of large stresses consistently lying in the inner shell rather than in the middle and outer shells.
For both types of aggregates, we also observe that the probability density functions of the rescaled internal stresses decay monotonically.
When we focus on stresses greater than half the maximum internal stress, we observe that for IAA-type aggregates, the inner shell has 4\% of such stresses in contrast to the middle shell that exhibits less than 1\% of such stresses, and to the outer shell where no stress greater than half the maximum stress is present.
In CCA-type aggregates, the inner shell contains roughly 8\% of stresses greater than half the maximum internal stress found across samples, the middle shell 3\%, while the outer shell has less than 1\% of such stresses. 
We also note that the stresses induced by a shear flow are much larger in CCA-type aggregates, with the largest internal stress found across samples being almost a factor of two greater than in the IAA case.
While it could not be ascertained whether a power-law or exponential decay behavior is observed in the middle and outer shells (partially due to the sparsity of the data), a power-law-like decay behavior is observed in the inner shell.

\begin{figure}
    \centering
    \includegraphics[width=0.45\linewidth]{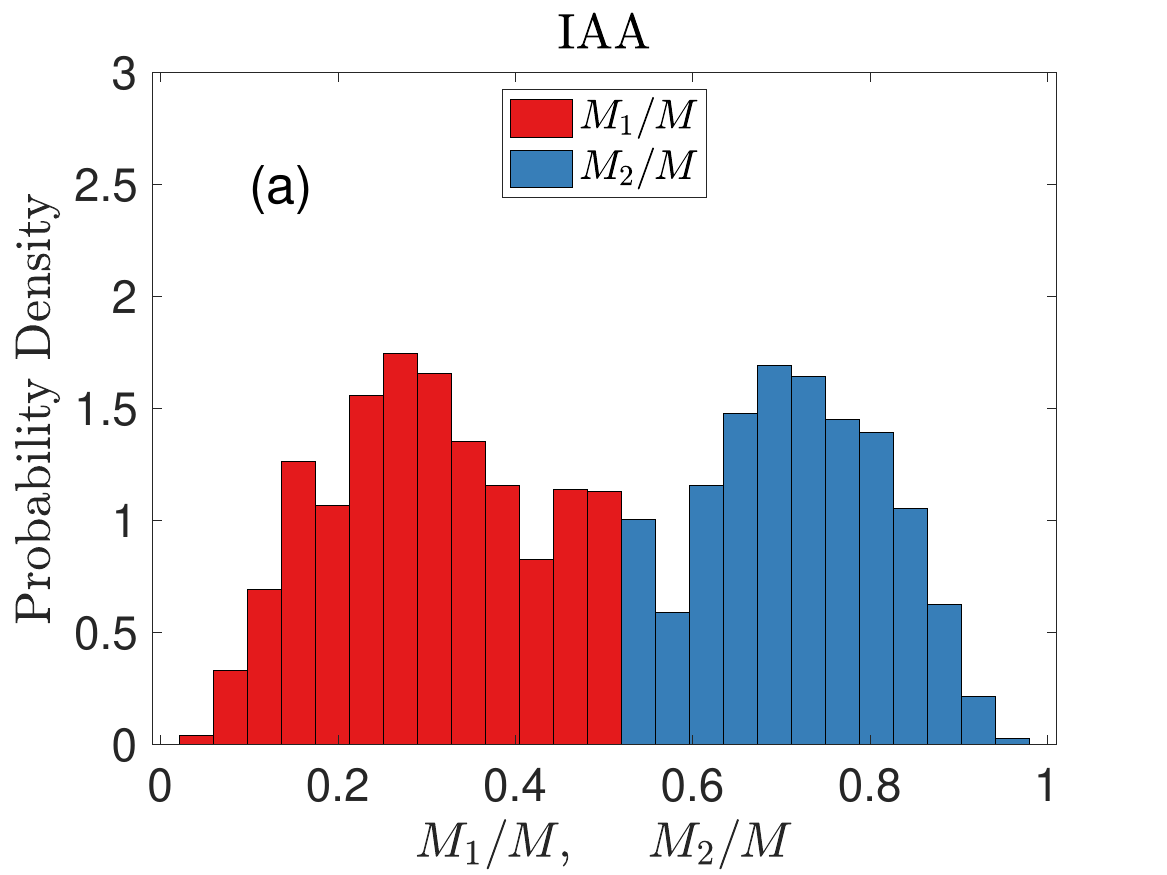}
    \includegraphics[width=0.45\linewidth]{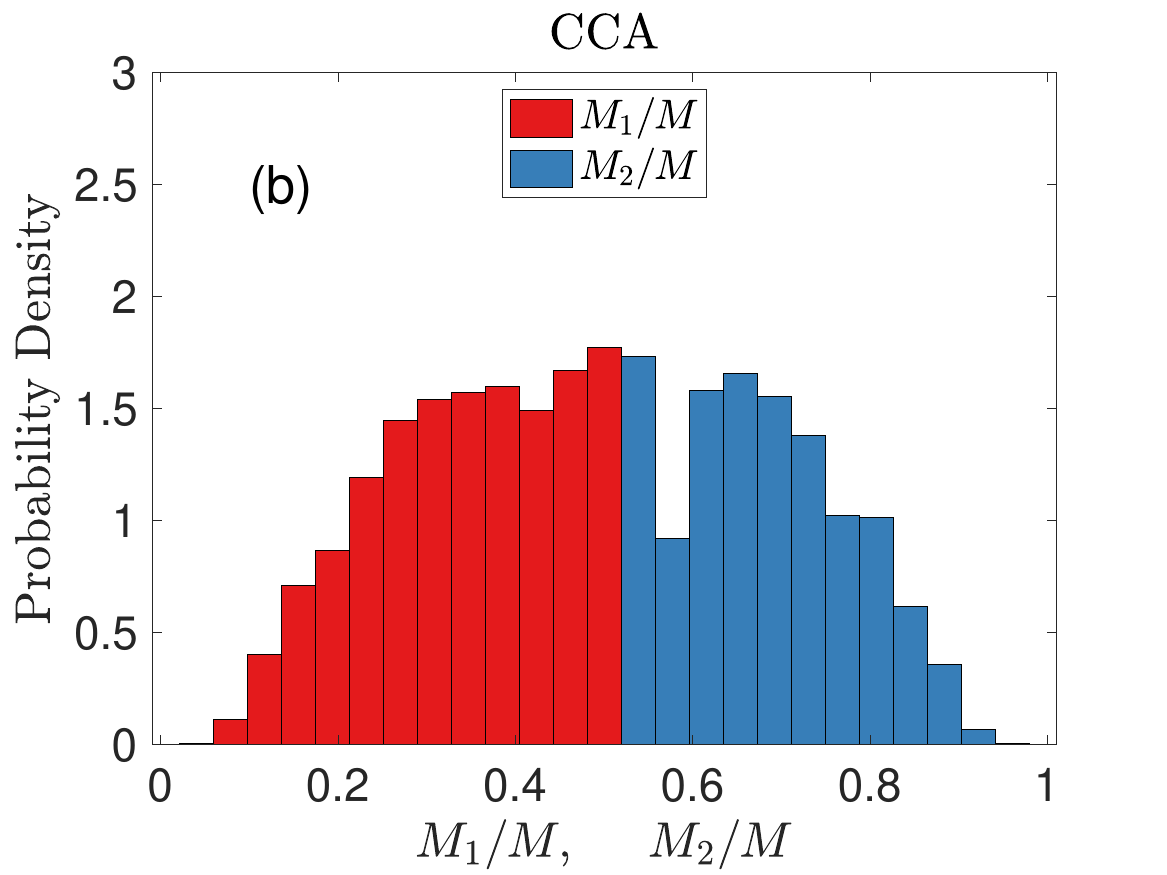}
    \caption{\label{fig:breakup_shear}
        Distribution of the relative masses of post-rupture aggregates for IAA-type, panel~(a), and CCA-type aggregates, panel~(b).
        Samples collected for aggregates with all reported $M$ values were used.
        We display the distribution of the ratio between the mass of the smaller of the two aggregates and the original, $M_1/M$, in red and that of the other ratio $M_2/M$ in blue.
        Results from aggregates that did not break up into two sub-aggregates were discarded.}
\end{figure}

The propensity of the largest stress caused by a shear flow to be located in the inner shell is reflected in the relative masses found after rupture, as shown in Figure~\ref{fig:breakup_shear}.
When breaking the aggregate into two at the location of the maximum internal stress, we see that broken-up aggregates are likely to have more evenly distributed masses (i.e., closer to half the initial aggregate mass) than in the Settling Case (see Figure~\ref{fig:breakup_settling}), and that it is rarer for them to have completely uneven masses.
We also observe that CCA-type aggregates are more likely to split evenly than their IAA counterparts, similarly to the trend observed in the Settling Case.

\begin{figure}
    \centering
    \includegraphics[width=\linewidth]{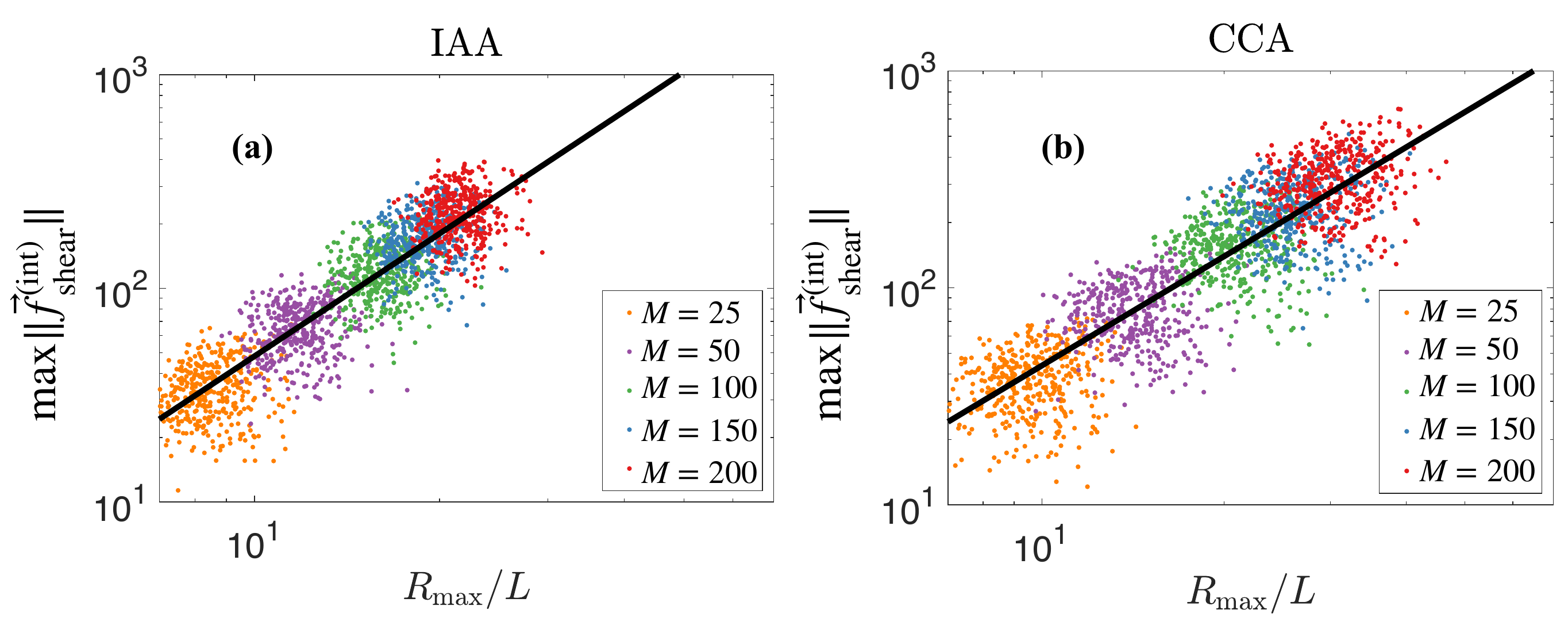}
    \caption{\label{fig:max_stress_vs_Rmax}
        Scatter plots of the maximum magnitude of the rescaled internal stress, $\vec{f}^{(\mathrm{int})}_\mathrm{shear}$, see \eqref{eq:fintshe} versus the maximum radius $R_{\mathrm{max}}$, see \eqref{eq:Rmax}, for the case of a shear background flow.
        All sample points are shown on a log-log scale.
        The number of cubes in aggregates span{s} from $M=25$ to $M=200$.
        The black lines represent the best linear fits to the data.}
\end{figure}

In Figure~\ref{fig:max_stress_vs_Rmax}, we show scatter plots of the maximum magnitude of the rescaled maximum internal stress $\max\left\lVert\vec{f}^{(\mathrm{int})}_\mathrm{shear}\right\rVert$ versus the maximum radius $R_\mathrm{max}$, see \eqref{eq:Rmax}, in a log-log scale, and perform linear regression.
We see that the maximum internal stress scales roughly quadratically with the maximum radius, both for IAA-type and CCA-type aggregates, as shown by the black lines in panels (a) and (b), the slopes of which are found to be 1.92 and 1.67, respectively.
As could be noticed in the distribution of all the internal stresses induced by a shear flow, see Figure~\ref{fig:shear_shells}, we observe that CCA-type aggregates experience much larger maximum internal stresses under shear than IAA-type aggregates, which is consistent with CCA-type aggregates having a larger radius for the same mass.

\section{\label{sec:disagg_Conclusion_disagg}Discussion and Conclusions}

In this work, we implemented a boundary integral formulation of Stokes equations to characterize stresses induced on fractal aggregates by an external force or background flow. 
We studied the external and internal stresses in low-Reynolds-number marine aggregates that are either settling under gravity in a fluid at rest or exposed to a laminar shear flow in the absence of forces. 
We investigated the impact of these different conditions on the distribution of internal stresses in the aggregates, quantified how the largest internal stresses scale with the mass and size of the aggregates, and computed the relative mass distributions of post-ruptured aggregates.

When imposing an external force given by an aggregate's apparent weight (Settling Case), we found that the distribution of the resulting internal stresses was similar in both more compact (IAA) and more wispy (CCA) aggregates.
More specifically, for both types of aggregates, we observed a monotonic decay in the probability density functions with distance from the center of mass of all three shells.
Furthermore, while small stresses were found to appear across all shells, large stresses were found to be rare and to arise mostly in the inner and middle shells, rather than the outer shell.
This indicates that both IAA and CCA-type aggregates tend to experience large stresses away from their periphery when settling under gravity.
These results are in good qualitative agreement with the findings of Gastaldi and Vanni~\cite{GastaldiVanni2011} for aggregates made of spheres and settling under the effect of a constant force in an unbounded fluid at rest.
These authors found that, for a range of aggregate types whose fractal dimensions are consistent with our IAA and CCA aggregates, large internal stresses occur less frequently than smaller internal stresses, and that they systematically lie away from the far edges of the aggregates.
They provided a detailed distribution of stresses as a function of distance from the center of mass that is broadly consistent with our results, although they reported an exponential decay in the frequency of large stresses while our results seem to indicate a power-law-like behavior.
Though the slightly different modes of formation of the aggregates studied may account for this discrepancy, conclusively distinguishing between those two tail decay behaviors would likely require significantly more data than the 400 samples we collected as these events are, by their very nature, rare.
We also found that, for both types of aggregates, the maximum internal stress scales as the ratio of an aggregate's apparent weight to the area of the thinnest connection, in this case a single square, with a proportionality constant $\alpha\approx 0.075$.
In other words, for the Settling Case, we obtained
\begin{equation}\label{eq:maxf_settl}
    \max\left\lVert\vec{f}^{(\mathrm{int})}\right\rVert \approx \alpha \frac{(\rho_\mathrm{agg}-\rho)gMV}{(2L)^2}.
\end{equation}

When settling aggregates were allowed to break at the location of the maximum internal stress, we found that the relative masses of the newly formed objects distribute in a fairly uneven manner, and that aggregates of both types rarely break at the far edges of their structure. 
On average, IAA-type aggregates break up in two aggregates whose relative masses are roughly 85\% and 15\% of the original mass, while CCA-type aggregates break up in two aggregates whose relative masses are roughly 80\% and 20\% of the original mass.
This is consistent with the rarity of large stresses near an aggregate's edges for both IAA- and CCA-type aggregates. 
While the overall distribution of the relative masses is similar in both types of aggregates, it is noteworthy that CCA-type aggregates have a greater likelihood to break close to their center of mass, compared to IAA-type aggregates, and thus would sometimes, albeit rarely, yield aggregates whose masses are roughly equal.
This is also consistent with the mode of formation of CCA-type aggregates that allows roughly equal-sized clusters to become connected by a thin bound, a feature absent in IAA-type aggregates clusters, see Figure~\ref{fig:shells}.

In aggregates exposed to laminar shear in the absence of forces (Shear Case), we found that the largest stresses are much more likely to be found close to the center of mass than near the edges, with an even greater prevalence than in the Settling Case. 
The stress distribution was again found to be qualitatively similar in IAA and CCA aggregates across shells, with large stresses being rare overall, but more frequent away from the edges of the aggregates.
When allowed to break at the location of the maximum internal stress, we found that  the relative masses of the newly formed objects distributed more {evenly} than for settling aggregates.
On average, IAA-type aggregates break up into two aggregates whose relative masses are roughly 70\% and 30\% of the original mass, while CCA-type aggregates break up into two aggregates whose relative masses are roughly 67\% and 33\% of the original mass.
Thus, like the Settling Case, CCA-type aggregates breakups are more likely to result in aggregates of comparable masses. 

We found that in the Shear Case the maximum internal stress scales almost quadratically with the maximum radius of the aggregates.
More specifically, from data fitting, we obtained
\begin{equation}\label{eq:maxf_shear}
    \max \left\Vert \vec{f}^{(\mathrm{int})} \right \Vert 
    \sim \gamma_t \mu \left( \frac{R_{\mathrm{max}}}{L} \right)^\beta,
\end{equation} 
where the exponent $\beta$ was estimated to be $\beta=1.92$ for IAA and $\beta=1.67$ for CCA.
We note that any shear can be written as the sum of a rotation and a strain:
\begin{equation}\label{eq:shear_flow}
    \vec{u}^{\infty}(\vec{x}) = \frac{1}{2}\vec{\omega}\times{\vec{x}} + {\bar{\bar E}}\vec{x}.
\end{equation}
In our case one can set $\vec{\omega}=-\gamma_t\hat{k}$, and $E_{1,2} = E_{2,1} = \gamma_t/2$ and all other entries of  ${\bar{\bar E}}$ to zero.
Using \eqref{eq:shear_flow} we thus recover the background flow $\vec{u}^{\infty}(\vec{x}) = {[\gamma_ty},0,0]^T$.
It was shown in Ref.~\cite{eunji} that extensional flows cause a straining force that scales quadratically with the characteristic length scale of fractal aggregates while torques scale as the length scale's third power. 
Thus, our findings indicate that the extensional contribution to the shear flow is what induces the internal stresses on the aggregates, and suggest that solid objects exposed to a low-Reynolds-number rotational background flow would simply rotate freely with the flow, without experiencing any additional internal stresses.
The larger stresses experienced by CCA-type aggregates appear to stem from the fact that these aggregates have a larger radius than IAA-type aggregates, which allows for a greater impact of the extensional portion of the flow.

\begin{figure}
    \centering
    \includegraphics[width=0.45\linewidth]{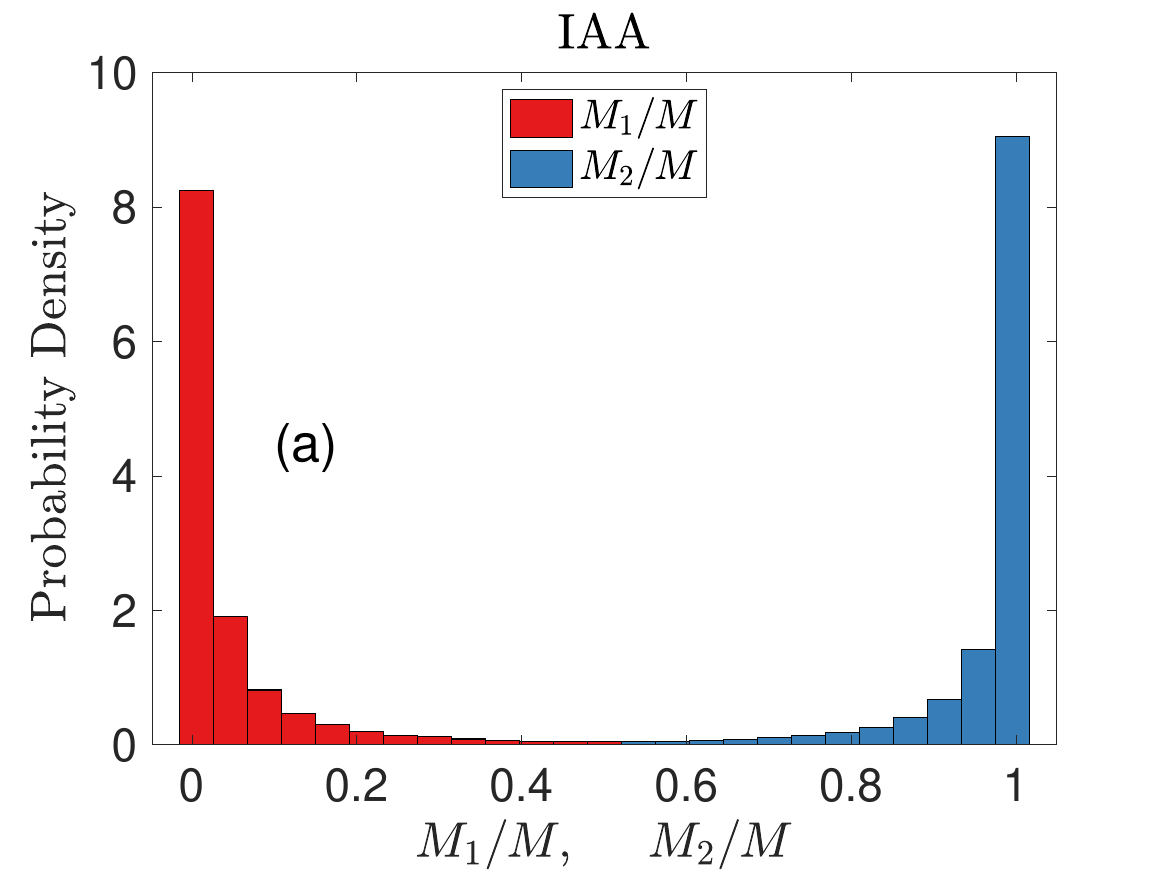}
    \includegraphics[width=0.45\linewidth]{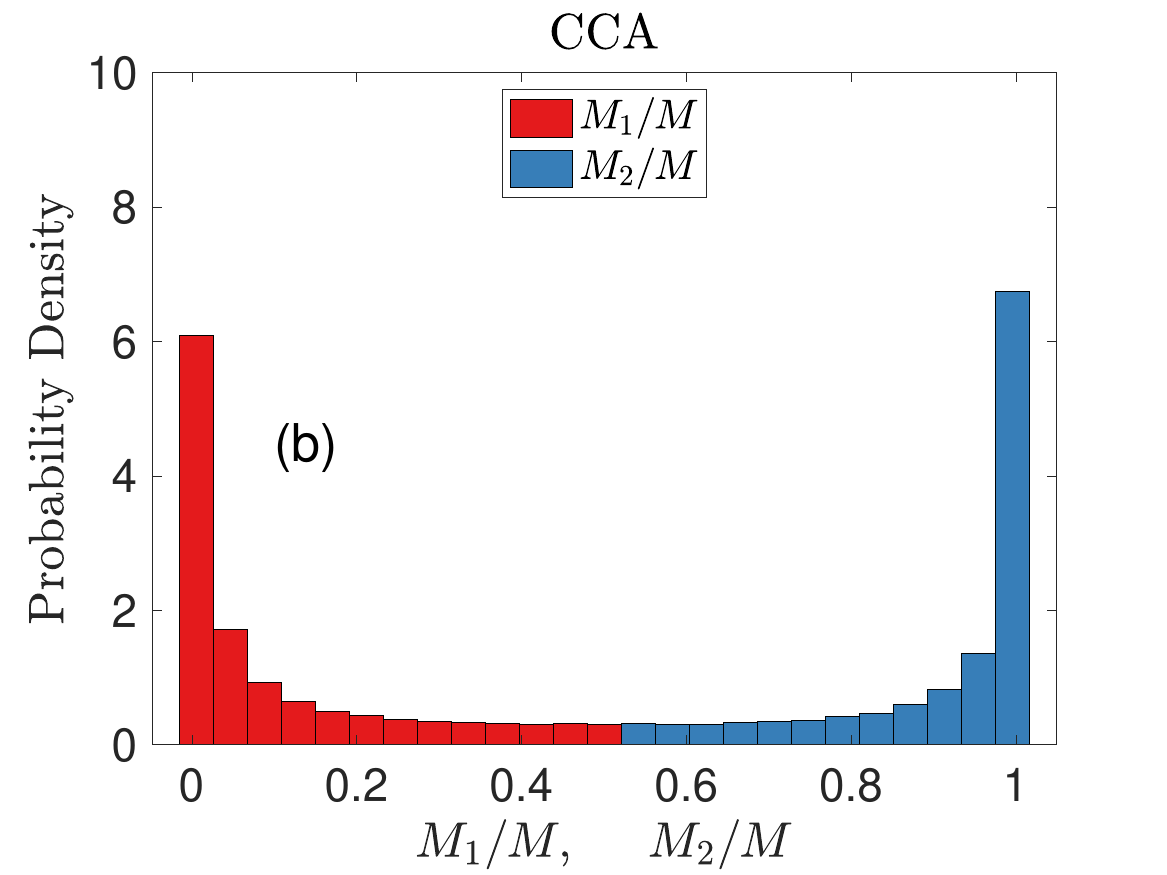}
    \caption{\label{fig:breakup_random}
        Relative mass distributions for IAA-type, panel~(a), and CCA-type aggregates, panel~(b), when every internal face of an aggregate are chosen in turn as a breakup location for all aggregate samples used.
        We display the distribution of the ratio between the mass of the smaller of the two aggregates and the original, $M_1/M$, in red and that of the other ratio $M_2/M$ in blue. Results from aggregates that did not break up into two sub-aggregates were discarded.}
\end{figure}

Our study provides useful insights on how to build a more accurate dynamical model of aggregation.
For instance, when information on the internal stress distribution is needed to randomly sample the location of rupture in a certain dynamical model, our findings indicate how internal stresses should be distributed based on their distance {to} the center of mass of the aggregate.
Traditionally, dynamical models of aggregation follow a reversible-aggregation approach in which aggregates break at a random location sampled with uniform probability~\cite{Kolb_1986}.
To compare the latter approach with our results, we computed the relative mass distribution of broken aggregates when aggregates are allowed to break into two at any one of their internal faces, selected in turn.
We note that, given the structure of IAA- and CCA-type aggregates~\cite{eunji}, there are many internal faces where severing a bond does not result in breakup because of a looped structure or adjacent connected faces.
Relative to all possible bonds, these cases arise for approximately 40\% of the internal faces in IAA-type aggregates and roughly 33\% in CCA-type aggregates.
Discarding those faces, we find, as shown in Figure~\ref{fig:breakup_random}, a distribution of the relative masses significantly different from both the Settling Case (Figure~\ref{fig:breakup_settling}) and the Shear Case (Figure~\ref{fig:breakup_shear}).
In fact, the location of the peaks in the relative mass distributions shown in Figure~\ref{fig:breakup_random} indicates that selecting the breakup location with uniform probability is more likely to break the original aggregates very unevenly, with typically just a few cubes detaching from the original aggregate both for IAA-type and CCA-type aggregates.
For IAA-type aggregates, the post-breakup masses were approximately 4\% and 96\% of the original mass, and for CCA-type aggregates, they were are approximately 10\% and 90\% of the original mass.
Therefore, selecting the breakup location from a uniform distribution gives far too much weight to peripheral locations, and a more accurate distribution of stresses that accounts for the structure of the aggregates is essential to the development of a physically relevant disaggregation model.

The results presented in this study show how internal stresses within low-Reynolds-number fractal aggregates distribute based on distance to the center of mass and suggest how this information can be directly used to build a dynamical model in which aggregates break at a location where they are more likely to experience large stresses.
For a given density difference, settling aggregates would tend toward an equilibrium size below which they would grow by aggregation and above which the stresses induced by their settling would cause them to break.
In the presence of a background flow, as often studied in experimental settings~\cite{SongRau2022, Song}, properly accounting for the extensional component of the flow is paramount to estimating the stresses acting on an aggregate.
This is because our results indicate that extension is dominant over the translational and rotational components of the flow when it comes to inducing internal stresses.
We expect that CCA-type objects would tend to break into smaller aggregates that would then grow into structures that more closely resemble IAA-type aggregates.

Ultimately, providing a well-justified disaggregation mechanism is an important step toward obtaining a more complete, and physically relevant model of aggregation that extends beyond the early stages of formation.
One potentially impactful feature that has yet to be properly characterized at low Reynolds number is the interactions between aggregates as they approach each other.
The extension of our boundary-integral approach to include hydrodynamic interactions between aggregates could further extend our understanding of aggregate dynamics.
A realistic dynamic model of aggregation should also include the possibility of aggregates deforming and restructuring.
In fact, deformation occurs in marine aggregates and is even more prevalent in other types of aggregates, such as granular aggregates, which have a tendency to deform rather than break~\cite{MuqtadirArgamAlDughaimiSaudDvorkin2020}.
The external and internal stresses discussed here could then be taken into account to more accurately characterize not only aggregate growth and breakup but also the evolution of their internal structure.

\begin{acknowledgments}
The authors acknowledge the support of the National Science Foundation Grant No.\ DMS-1840265.
\end{acknowledgments}

\bibliography{refs}

\begin{thebibliography}{26}%
\makeatletter
\providecommand \@ifxundefined [1]{%
 \@ifx{#1\undefined}
}%
\providecommand \@ifnum [1]{%
 \ifnum #1\expandafter \@firstoftwo
 \else \expandafter \@secondoftwo
 \fi
}%
\providecommand \@ifx [1]{%
 \ifx #1\expandafter \@firstoftwo
 \else \expandafter \@secondoftwo
 \fi
}%
\providecommand \natexlab [1]{#1}%
\providecommand \enquote  [1]{``#1''}%
\providecommand \bibnamefont  [1]{#1}%
\providecommand \bibfnamefont [1]{#1}%
\providecommand \citenamefont [1]{#1}%
\providecommand \href@noop [0]{\@secondoftwo}%
\providecommand \href [0]{\begingroup \@sanitize@url \@href}%
\providecommand \@href[1]{\@@startlink{#1}\@@href}%
\providecommand \@@href[1]{\endgroup#1\@@endlink}%
\providecommand \@sanitize@url [0]{\catcode `\\12\catcode `\$12\catcode `\&12\catcode `\#12\catcode `\^12\catcode `\_12\catcode `\%12\relax}%
\providecommand \@@startlink[1]{}%
\providecommand \@@endlink[0]{}%
\providecommand \url  [0]{\begingroup\@sanitize@url \@url }%
\providecommand \@url [1]{\endgroup\@href {#1}{\urlprefix }}%
\providecommand \urlprefix  [0]{URL }%
\providecommand \Eprint [0]{\href }%
\providecommand \doibase [0]{https://doi.org/}%
\providecommand \selectlanguage [0]{\@gobble}%
\providecommand \bibinfo  [0]{\@secondoftwo}%
\providecommand \bibfield  [0]{\@secondoftwo}%
\providecommand \translation [1]{[#1]}%
\providecommand \BibitemOpen [0]{}%
\providecommand \bibitemStop [0]{}%
\providecommand \bibitemNoStop [0]{.\EOS\space}%
\providecommand \EOS [0]{\spacefactor3000\relax}%
\providecommand \BibitemShut  [1]{\csname bibitem#1\endcsname}%
\let\auto@bib@innerbib\@empty
\bibitem [{\citenamefont {Burd}\ and\ \citenamefont {Jackson}(2009)}]{burd_particle_2009}%
  \BibitemOpen
  \bibfield  {author} {\bibinfo {author} {\bibfnamefont {A.~B.}\ \bibnamefont {Burd}}\ and\ \bibinfo {author} {\bibfnamefont {G.~A.}\ \bibnamefont {Jackson}},\ }\bibfield  {title} {\bibinfo {title} {Particle aggregation},\ }\href {https://doi.org/10.1146/annurev.marine.010908.163904} {\bibfield  {journal} {\bibinfo  {journal} {Ann.\ Rev.\ Mar.\ Sci.}\ }\textbf {\bibinfo {volume} {1}},\ \bibinfo {pages} {65} (\bibinfo {year} {2009})}\BibitemShut {NoStop}%
\bibitem [{\citenamefont {Honjo}\ \emph {et~al.}(2014)\citenamefont {Honjo}, \citenamefont {Eglinton}, \citenamefont {Taylor}, \citenamefont {Ulmer}, \citenamefont {Sievert}, \citenamefont {Bracher}, \citenamefont {German}, \citenamefont {Edgcomb}, \citenamefont {Fran\c{c}ois}, \citenamefont {Iglesias-Rodriguez}, \citenamefont {Van~Mooy},\ and\ \citenamefont {Rapeta}}]{carbon}%
  \BibitemOpen
  \bibfield  {author} {\bibinfo {author} {\bibfnamefont {S.}~\bibnamefont {Honjo}}, \bibinfo {author} {\bibfnamefont {T.}~\bibnamefont {Eglinton}}, \bibinfo {author} {\bibfnamefont {C.}~\bibnamefont {Taylor}}, \bibinfo {author} {\bibfnamefont {K.}~\bibnamefont {Ulmer}}, \bibinfo {author} {\bibfnamefont {S.}~\bibnamefont {Sievert}}, \bibinfo {author} {\bibfnamefont {A.}~\bibnamefont {Bracher}}, \bibinfo {author} {\bibfnamefont {C.}~\bibnamefont {German}}, \bibinfo {author} {\bibfnamefont {V.}~\bibnamefont {Edgcomb}}, \bibinfo {author} {\bibfnamefont {R.}~\bibnamefont {Fran\c{c}ois}}, \bibinfo {author} {\bibfnamefont {M.}~\bibnamefont {Iglesias-Rodriguez}}, \bibinfo {author} {\bibfnamefont {B.}~\bibnamefont {Van~Mooy}},\ and\ \bibinfo {author} {\bibfnamefont {D.}~\bibnamefont {Rapeta}},\ }\bibfield  {title} {\bibinfo {title} {Understanding the role of the biological pump in the global carbon cycle: An imperative for ocean science},\ }\href {https://doi.org/10.5670/oceanog.2014.78} {\bibfield  {journal}
  {\bibinfo  {journal} {Oceanography}\ }\textbf {\bibinfo {volume} {27}},\ \bibinfo {pages} {10} (\bibinfo {year} {2014})}\BibitemShut {NoStop}%
\bibitem [{\citenamefont {Coyle}\ \emph {et~al.}(2020)\citenamefont {Coyle}, \citenamefont {Hardiman},\ and\ \citenamefont {Driscoll}}]{microplastics}%
  \BibitemOpen
  \bibfield  {author} {\bibinfo {author} {\bibfnamefont {R.}~\bibnamefont {Coyle}}, \bibinfo {author} {\bibfnamefont {G.}~\bibnamefont {Hardiman}},\ and\ \bibinfo {author} {\bibfnamefont {K.~O.}\ \bibnamefont {Driscoll}},\ }\bibfield  {title} {\bibinfo {title} {Microplastics in the marine environment: A review of their sources, distribution processes, uptake and exchange in ecosystems},\ }\href {https://doi.org/10.1016/j.cscee.2020.100010} {\bibfield  {journal} {\bibinfo  {journal} {Case Stud.\ Chem.\ Environ.\ Eng.}\ }\textbf {\bibinfo {volume} {2}},\ \bibinfo {pages} {100010} (\bibinfo {year} {2020})}\BibitemShut {NoStop}%
\bibitem [{\citenamefont {Meakin}(1984)}]{meakin_3D_paper}%
  \BibitemOpen
  \bibfield  {author} {\bibinfo {author} {\bibfnamefont {P.}~\bibnamefont {Meakin}},\ }\bibfield  {title} {\bibinfo {title} {Diffusion-limited aggregation in three dimensions: Results from a new cluster-cluster aggregation model},\ }\href {https://doi.org/10.1016/0021-9797(84)90252-2} {\bibfield  {journal} {\bibinfo  {journal} {J.\ Colloid Interface Sci.}\ }\textbf {\bibinfo {volume} {102}},\ \bibinfo {pages} {491} (\bibinfo {year} {1984})}\BibitemShut {NoStop}%
\bibitem [{\citenamefont {Jungblut}\ \emph {et~al.}(2019)\citenamefont {Jungblut}, \citenamefont {Joswig},\ and\ \citenamefont {Eychm{\"u}ller}}]{jung}%
  \BibitemOpen
  \bibfield  {author} {\bibinfo {author} {\bibfnamefont {S.}~\bibnamefont {Jungblut}}, \bibinfo {author} {\bibfnamefont {J.-O.}\ \bibnamefont {Joswig}},\ and\ \bibinfo {author} {\bibfnamefont {A.}~\bibnamefont {Eychm{\"u}ller}},\ }\bibfield  {title} {\bibinfo {title} {Diffusion-limited cluster aggregation: Impact of rotational diffusion},\ }\href {https://doi.org/10.1021/acs.jpcc.8b10805} {\bibfield  {journal} {\bibinfo  {journal} {J.\ Phys.\ Chem.\ C}\ }\textbf {\bibinfo {volume} {123}},\ \bibinfo {pages} {950} (\bibinfo {year} {2019})}\BibitemShut {NoStop}%
\bibitem [{\citenamefont {Polimeno}\ \emph {et~al.}(2022)\citenamefont {Polimeno}, \citenamefont {Kim},\ and\ \citenamefont {Blanchette}}]{Polimeno2022}%
  \BibitemOpen
  \bibfield  {author} {\bibinfo {author} {\bibfnamefont {M.}~\bibnamefont {Polimeno}}, \bibinfo {author} {\bibfnamefont {C.}~\bibnamefont {Kim}},\ and\ \bibinfo {author} {\bibfnamefont {F.}~\bibnamefont {Blanchette}},\ }\bibfield  {title} {\bibinfo {title} {Toward a realistic model of diffusion-limited aggregation: Rotation, size-dependent diffusivities, and settling},\ }\href {https://doi.org/10.1021/acsomega.2c03547} {\bibfield  {journal} {\bibinfo  {journal} {ACS Omega}\ }\textbf {\bibinfo {volume} {7}},\ \bibinfo {pages} {40826} (\bibinfo {year} {2022})}\BibitemShut {NoStop}%
\bibitem [{\citenamefont {Song}\ and\ \citenamefont {Rau}(2020)}]{Song}%
  \BibitemOpen
  \bibfield  {author} {\bibinfo {author} {\bibfnamefont {Y.}~\bibnamefont {Song}}\ and\ \bibinfo {author} {\bibfnamefont {M.~J.}\ \bibnamefont {Rau}},\ }\bibfield  {title} {\bibinfo {title} {Viscous fluid flow inside an oscillating cylinder and its extension to {S}tokes’ second problem},\ }\href {https://doi.org/10.1063/1.5144415} {\bibfield  {journal} {\bibinfo  {journal} {Phys.\ Fluids}\ }\textbf {\bibinfo {volume} {32}},\ \bibinfo {pages} {043601} (\bibinfo {year} {2020})}\BibitemShut {NoStop}%
\bibitem [{\citenamefont {De~La Rosa~Zambrano}\ \emph {et~al.}(2018)\citenamefont {De~La Rosa~Zambrano}, \citenamefont {Verhille},\ and\ \citenamefont {Le~Gal}}]{DeLaRosaZambranVerhilleLeGal2018}%
  \BibitemOpen
  \bibfield  {author} {\bibinfo {author} {\bibfnamefont {H.~M.}\ \bibnamefont {De~La Rosa~Zambrano}}, \bibinfo {author} {\bibfnamefont {G.}~\bibnamefont {Verhille}},\ and\ \bibinfo {author} {\bibfnamefont {P.}~\bibnamefont {Le~Gal}},\ }\bibfield  {title} {\bibinfo {title} {Fragmentation of magnetic particle aggregates in turbulence},\ }\href {https://doi.org/10.1103/PhysRevFluids.3.084605} {\bibfield  {journal} {\bibinfo  {journal} {Phys.\ Rev.\ Fluids}\ }\textbf {\bibinfo {volume} {3}},\ \bibinfo {pages} {084605} (\bibinfo {year} {2018})}\BibitemShut {NoStop}%
\bibitem [{\citenamefont {Song}\ and\ \citenamefont {Rau}(2022)}]{SongRau2022}%
  \BibitemOpen
  \bibfield  {author} {\bibinfo {author} {\bibfnamefont {Y.}~\bibnamefont {Song}}\ and\ \bibinfo {author} {\bibfnamefont {M.~J.}\ \bibnamefont {Rau}},\ }\bibfield  {title} {\bibinfo {title} {A novel method to study the fragmentation behavior of marine snow aggregates in controlled shear flow},\ }\href {https://doi.org/10.1002/lom3.10509} {\bibfield  {journal} {\bibinfo  {journal} {Limnol.\ Oceanogr.: Methods}\ }\textbf {\bibinfo {volume} {20}},\ \bibinfo {pages} {618} (\bibinfo {year} {2022})}\BibitemShut {NoStop}%
\bibitem [{\citenamefont {Brouzet}\ \emph {et~al.}(2021)\citenamefont {Brouzet}, \citenamefont {Guin{\'e}}, \citenamefont {Dalbe}, \citenamefont {Favier}, \citenamefont {Vandenberghe}, \citenamefont {Villermaux},\ and\ \citenamefont {Verhille}}]{BrouzetGuineDalbeFavierVandenbergheVillermauxVerhille2021}%
  \BibitemOpen
  \bibfield  {author} {\bibinfo {author} {\bibfnamefont {C.}~\bibnamefont {Brouzet}}, \bibinfo {author} {\bibfnamefont {R.}~\bibnamefont {Guin{\'e}}}, \bibinfo {author} {\bibfnamefont {M.-J.}\ \bibnamefont {Dalbe}}, \bibinfo {author} {\bibfnamefont {B.}~\bibnamefont {Favier}}, \bibinfo {author} {\bibfnamefont {N.}~\bibnamefont {Vandenberghe}}, \bibinfo {author} {\bibfnamefont {E.}~\bibnamefont {Villermaux}},\ and\ \bibinfo {author} {\bibfnamefont {G.}~\bibnamefont {Verhille}},\ }\bibfield  {title} {\bibinfo {title} {Laboratory model for plastic fragmentation in the turbulent ocean},\ }\href {https://doi.org/10.1103/PhysRevFluids.6.024601} {\bibfield  {journal} {\bibinfo  {journal} {Phys.\ Rev.\ Fluids}\ }\textbf {\bibinfo {volume} {6}},\ \bibinfo {pages} {024601} (\bibinfo {year} {2021})}\BibitemShut {NoStop}%
\bibitem [{\citenamefont {Frungieri}\ and\ \citenamefont {Vanni}(2021)}]{FRUNGIERI2021357}%
  \BibitemOpen
  \bibfield  {author} {\bibinfo {author} {\bibfnamefont {G.}~\bibnamefont {Frungieri}}\ and\ \bibinfo {author} {\bibfnamefont {M.}~\bibnamefont {Vanni}},\ }\bibfield  {title} {\bibinfo {title} {Aggregation and breakup of colloidal particle aggregates in shear flow: A combined {M}onte {C}arlo--{S}tokesian dynamics approach},\ }\href {https://doi.org/10.1016/j.powtec.2021.04.076} {\bibfield  {journal} {\bibinfo  {journal} {Powder Technol.}\ }\textbf {\bibinfo {volume} {388}},\ \bibinfo {pages} {357} (\bibinfo {year} {2021})}\BibitemShut {NoStop}%
\bibitem [{\citenamefont {Zhao}\ \emph {et~al.}(2023)\citenamefont {Zhao}, \citenamefont {Vowinckel}, \citenamefont {Hsu}, \citenamefont {Bai},\ and\ \citenamefont {Meiburg}}]{ZhaoVowinckelHsuBaiMeiburg2023}%
  \BibitemOpen
  \bibfield  {author} {\bibinfo {author} {\bibfnamefont {K.}~\bibnamefont {Zhao}}, \bibinfo {author} {\bibfnamefont {B.}~\bibnamefont {Vowinckel}}, \bibinfo {author} {\bibfnamefont {T.-J.}\ \bibnamefont {Hsu}}, \bibinfo {author} {\bibfnamefont {B.}~\bibnamefont {Bai}},\ and\ \bibinfo {author} {\bibfnamefont {E.}~\bibnamefont {Meiburg}},\ }\bibfield  {title} {\bibinfo {title} {Cohesive sediment: Intermediate shear produces maximum aggregate size},\ }\href {https://doi.org/10.1017/jfm.2023.380} {\bibfield  {journal} {\bibinfo  {journal} {J.\ Fluid Mech.}\ }\textbf {\bibinfo {volume} {965}},\ \bibinfo {pages} {A5} (\bibinfo {year} {2023})}\BibitemShut {NoStop}%
\bibitem [{\citenamefont {Kolb}(1986)}]{Kolb_1986}%
  \BibitemOpen
  \bibfield  {author} {\bibinfo {author} {\bibfnamefont {M.}~\bibnamefont {Kolb}},\ }\bibfield  {title} {\bibinfo {title} {Reversible diffusion-limited cluster aggregation},\ }\href {https://doi.org/10.1088/0305-4470/19/5/009} {\bibfield  {journal} {\bibinfo  {journal} {J.\ Phys.\ A: Math.\ Gen.}\ }\textbf {\bibinfo {volume} {19}},\ \bibinfo {pages} {L263} (\bibinfo {year} {1986})}\BibitemShut {NoStop}%
\bibitem [{\citenamefont {Zaccone}\ \emph {et~al.}(2009)\citenamefont {Zaccone}, \citenamefont {Soos}, \citenamefont {Lattuada}, \citenamefont {Wu}, \citenamefont {B{\"a}bler},\ and\ \citenamefont {Morbidelli}}]{ZacconeSoosLattuadaWuBablerMatthausMorbidelli2009}%
  \BibitemOpen
  \bibfield  {author} {\bibinfo {author} {\bibfnamefont {A.}~\bibnamefont {Zaccone}}, \bibinfo {author} {\bibfnamefont {M.}~\bibnamefont {Soos}}, \bibinfo {author} {\bibfnamefont {M.}~\bibnamefont {Lattuada}}, \bibinfo {author} {\bibfnamefont {H.}~\bibnamefont {Wu}}, \bibinfo {author} {\bibfnamefont {M.~U.}\ \bibnamefont {B{\"a}bler}},\ and\ \bibinfo {author} {\bibfnamefont {M.}~\bibnamefont {Morbidelli}},\ }\bibfield  {title} {\bibinfo {title} {Breakup of dense colloidal aggregates under hydrodynamic stresses},\ }\href {https://doi.org/10.1103/PhysRevE.79.061401} {\bibfield  {journal} {\bibinfo  {journal} {Phys.\ Rev.\ E}\ }\textbf {\bibinfo {volume} {79}},\ \bibinfo {pages} {061401} (\bibinfo {year} {2009})}\BibitemShut {NoStop}%
\bibitem [{\citenamefont {Gastaldi}\ and\ \citenamefont {Vanni}(2011)}]{GastaldiVanni2011}%
  \BibitemOpen
  \bibfield  {author} {\bibinfo {author} {\bibfnamefont {A.}~\bibnamefont {Gastaldi}}\ and\ \bibinfo {author} {\bibfnamefont {M.}~\bibnamefont {Vanni}},\ }\bibfield  {title} {\bibinfo {title} {The distribution of stresses in rigid fractal-like aggregates in a uniform flow field},\ }\href {https://doi.org/10.1016/j.jcis.2011.01.080} {\bibfield  {journal} {\bibinfo  {journal} {J.\ Colloid Interface Sci.}\ }\textbf {\bibinfo {volume} {357}},\ \bibinfo {pages} {18} (\bibinfo {year} {2011})}\BibitemShut {NoStop}%
\bibitem [{\citenamefont {Happel}\ and\ \citenamefont {Brenner}(1983)}]{happel1965low}%
  \BibitemOpen
  \bibfield  {author} {\bibinfo {author} {\bibfnamefont {J.}~\bibnamefont {Happel}}\ and\ \bibinfo {author} {\bibfnamefont {H.}~\bibnamefont {Brenner}},\ }\href {https://doi.org/10.1007/978-94-009-8352-6} {\emph {\bibinfo {title} {Low Reynolds Number Hydrodynamics: With Special Applications to Particulate Media}}}\ (\bibinfo  {publisher} {Springer},\ \bibinfo {address} {Dordrecht},\ \bibinfo {year} {1983})\BibitemShut {NoStop}%
\bibitem [{\citenamefont {German}(2014)}]{German}%
  \BibitemOpen
  \bibfield  {author} {\bibinfo {author} {\bibfnamefont {R.~M.}\ \bibnamefont {German}},\ }\bibfield  {title} {\bibinfo {title} {Coordination number changes during powder densification},\ }\href {https://doi.org/10.1016/j.powtec.2013.12.006} {\bibfield  {journal} {\bibinfo  {journal} {Powder Technol.}\ }\textbf {\bibinfo {volume} {253}},\ \bibinfo {pages} {368} (\bibinfo {year} {2014})}\BibitemShut {NoStop}%
\bibitem [{\citenamefont {Pozrikidis}(1992)}]{pozrikidis_1992}%
  \BibitemOpen
  \bibfield  {author} {\bibinfo {author} {\bibfnamefont {C.}~\bibnamefont {Pozrikidis}},\ }\href {https://doi.org/10.1017/CBO9780511624124} {\emph {\bibinfo {title} {Boundary Integral and Singularity Methods for Linearized Viscous Flow}}}\ (\bibinfo  {publisher} {Cambridge University Press},\ \bibinfo {year} {1992})\BibitemShut {NoStop}%
\bibitem [{\citenamefont {Oman}\ and\ \citenamefont {Oman}(2015)}]{Oman}%
  \BibitemOpen
  \bibfield  {author} {\bibinfo {author} {\bibfnamefont {D.}~\bibnamefont {Oman}}\ and\ \bibinfo {author} {\bibfnamefont {R.}~\bibnamefont {Oman}},\ }\href@noop {} {\emph {\bibinfo {title} {How to Solve Physics Problems}}}\ (\bibinfo  {publisher} {McGraw Hill},\ \bibinfo {year} {2015})\BibitemShut {NoStop}%
\bibitem [{\citenamefont {Witten}\ and\ \citenamefont {Sander}(1981)}]{witten1981}%
  \BibitemOpen
  \bibfield  {author} {\bibinfo {author} {\bibfnamefont {T.~A.}\ \bibnamefont {Witten}}\ and\ \bibinfo {author} {\bibfnamefont {L.~M.}\ \bibnamefont {Sander}},\ }\bibfield  {title} {\bibinfo {title} {Diffusion-limited aggregation, a kinetic critical phenomenon},\ }\href {https://doi.org/10.1103/PhysRevLett.47.1400} {\bibfield  {journal} {\bibinfo  {journal} {Phys.\ Rev.\ Lett.}\ }\textbf {\bibinfo {volume} {47}},\ \bibinfo {pages} {1400} (\bibinfo {year} {1981})}\BibitemShut {NoStop}%
\bibitem [{\citenamefont {Meakin}(1983)}]{meakin1983}%
  \BibitemOpen
  \bibfield  {author} {\bibinfo {author} {\bibfnamefont {P.}~\bibnamefont {Meakin}},\ }\bibfield  {title} {\bibinfo {title} {Diffusion-controlled cluster formation in 2--6-dimensional space},\ }\href {https://doi.org/10.1103/PhysRevA.27.1495} {\bibfield  {journal} {\bibinfo  {journal} {Phys.\ Rev.\ A}\ }\textbf {\bibinfo {volume} {27}},\ \bibinfo {pages} {1495} (\bibinfo {year} {1983})}\BibitemShut {NoStop}%
\bibitem [{\citenamefont {Yoo}\ \emph {et~al.}(2020)\citenamefont {Yoo}, \citenamefont {Khatri},\ and\ \citenamefont {Blanchette}}]{eunji}%
  \BibitemOpen
  \bibfield  {author} {\bibinfo {author} {\bibfnamefont {E.}~\bibnamefont {Yoo}}, \bibinfo {author} {\bibfnamefont {S.}~\bibnamefont {Khatri}},\ and\ \bibinfo {author} {\bibfnamefont {F.}~\bibnamefont {Blanchette}},\ }\bibfield  {title} {\bibinfo {title} {Hydrodynamic forces on randomly formed marine aggregates},\ }\href {https://doi.org/10.1103/PhysRevFluids.5.044305} {\bibfield  {journal} {\bibinfo  {journal} {Phys.\ Rev.\ Fluids}\ }\textbf {\bibinfo {volume} {5}},\ \bibinfo {pages} {044305} (\bibinfo {year} {2020})}\BibitemShut {NoStop}%
\bibitem [{\citenamefont {Finkel}\ \emph {et~al.}(2009)\citenamefont {Finkel}, \citenamefont {Beardall}, \citenamefont {Flynn}, \citenamefont {Quigg}, \citenamefont {Rees},\ and\ \citenamefont {Raven}}]{plankton}%
  \BibitemOpen
  \bibfield  {author} {\bibinfo {author} {\bibfnamefont {Z.~V.}\ \bibnamefont {Finkel}}, \bibinfo {author} {\bibfnamefont {J.}~\bibnamefont {Beardall}}, \bibinfo {author} {\bibfnamefont {K.~J.}\ \bibnamefont {Flynn}}, \bibinfo {author} {\bibfnamefont {A.}~\bibnamefont {Quigg}}, \bibinfo {author} {\bibfnamefont {T.~A.~V.}\ \bibnamefont {Rees}},\ and\ \bibinfo {author} {\bibfnamefont {J.~A.}\ \bibnamefont {Raven}},\ }\bibfield  {title} {\bibinfo {title} {Phytoplankton in a changing world: Cell size and elemental stoichiometry},\ }\href {https://doi.org/10.1093/plankt/fbp098} {\bibfield  {journal} {\bibinfo  {journal} {J.\ Plankton Res.}\ }\textbf {\bibinfo {volume} {32}},\ \bibinfo {pages} {119} (\bibinfo {year} {2009})}\BibitemShut {NoStop}%
\bibitem [{\citenamefont {Alldredge}\ and\ \citenamefont {Gotschalk}(1988)}]{marinesnow}%
  \BibitemOpen
  \bibfield  {author} {\bibinfo {author} {\bibfnamefont {A.~L.}\ \bibnamefont {Alldredge}}\ and\ \bibinfo {author} {\bibfnamefont {C.}~\bibnamefont {Gotschalk}},\ }\bibfield  {title} {\bibinfo {title} {In situ settling behavior of marine snow},\ }\href {https://doi.org/10.4319/lo.1988.33.3.0339} {\bibfield  {journal} {\bibinfo  {journal} {Limnol.\ Oceanogr.}\ }\textbf {\bibinfo {volume} {33}},\ \bibinfo {pages} {339} (\bibinfo {year} {1988})}\BibitemShut {NoStop}%
\bibitem [{\citenamefont {Bertsekas}(1982)}]{bertsekas2014constrained}%
  \BibitemOpen
  \bibfield  {author} {\bibinfo {author} {\bibfnamefont {D.~P.}\ \bibnamefont {Bertsekas}},\ }\href@noop {} {\emph {\bibinfo {title} {Constrained Optimization and Lagrange Multiplier Methods}}}\ (\bibinfo  {publisher} {Academic Press},\ \bibinfo {year} {1982})\BibitemShut {NoStop}%
\bibitem [{\citenamefont {Muqtadir}\ \emph {et~al.}(2020)\citenamefont {Muqtadir}, \citenamefont {Al-Dughaimi},\ and\ \citenamefont {Dvorkin}}]{MuqtadirArgamAlDughaimiSaudDvorkin2020}%
  \BibitemOpen
  \bibfield  {author} {\bibinfo {author} {\bibfnamefont {A.}~\bibnamefont {Muqtadir}}, \bibinfo {author} {\bibfnamefont {S.}~\bibnamefont {Al-Dughaimi}},\ and\ \bibinfo {author} {\bibfnamefont {J.}~\bibnamefont {Dvorkin}},\ }\bibfield  {title} {\bibinfo {title} {Deformation of granular aggregates: Static and dynamic bulk moduli},\ }\href {https://doi.org/10.1029/2019JB018604} {\bibfield  {journal} {\bibinfo  {journal} {J.\ Geophys.\ Res.\ Solid Earth}\ }\textbf {\bibinfo {volume} {125}},\ \bibinfo {pages} {e2019JB018604} (\bibinfo {year} {2020})}\BibitemShut {NoStop}%
\end{thebibliography}%

\end{document}